\documentclass[useAMS,usenatbib]{mn2e}
\usepackage{graphicx}
\usepackage{color}
\newcommand{\Msun}{M$_\odot$}
\newcommand{\Lsun}{L$_\odot$}
\newcommand{\Zsun}{Z$_\odot$}

\title[The effect of feedback and reionization on star formation in low-mass dwarf galaxy haloes]{The effect of feedback and reionization on star formation in low-mass dwarf galaxy haloes}
\author[C. M. Simpson et al.]{Christine M. Simpson$^{1}$\thanks{E-mail: csimpson@astro.columbia.edu}, Greg L. Bryan$^{1}$, Kathryn V. Johnston$^{1}$, 
\newauthor
Britton D. Smith$^{2}$, Mordecai-Mark Mac Low$^{3,1}$, Sanjib Sharma$^{4}$ and
\newauthor
Jason Tumlinson$^{5}$\\
$^{1}$Department of Astronomy \& Astrophysics, Columbia University, New York, NY 10027 USA\\
$^{2}$Department of Physics and Astronomy, Michigan State University, East Lansing, MI 48824 USA\\
$^{3}$Department of Astrophysics, American Museum of Natural History, New York, NY 10024 USA\\
$^{4}$Sydney Institute for Astronomy, School of Physics, University of Sydney, NSW 2006, Australia\\
$^{5}$Space Telescope Science Institute, 3700 San Martin Drive, Baltimore, MD, 21218 USA}
\begin{document}

\date{}

\pagerange{\pageref{firstpage}--\pageref{lastpage}} \pubyear{2012}

\maketitle

\label{firstpage}

\begin{abstract}

We simulate the evolution of a 10$^9$ \Msun\ dark matter halo in a cosmological setting with an adaptive-mesh refinement code as an analogue to local low luminosity dwarf irregular and dwarf spheroidal galaxies.  The primary goal of our study is to investigate the roles of reionization and supernova feedback in determining the star formation histories of low mass dwarf galaxies.  We include a wide range of physical effects, including metal cooling, molecular hydrogen formation and cooling, photoionization and photodissociation from a metagalactic (but not local) background, a simple prescription for self-shielding, star formation, and a simple model for supernova driven energetic feedback.  To better understand the impact of each physical effect, we carry out simulations excluding each major effect in turn.  We find that reionization is primarily responsible for expelling most of the gas in our simulations, but that supernova feedback is required to disperse the dense, cold gas in the core of the halo.  Moreover, we show that the timing of reionization can produce an order of magnitude difference in the final stellar mass of the system.  For our full physics run with reionization at $z=9$, we find a stellar mass of about 10$^5$ \Msun\ at $z=0$, and a mass-to-light ratio within the half-light radius of approximately 130 \Msun/\Lsun, consistent with observed low-luminosity dwarfs.  However, the resulting median stellar metallicity is 0.06 \Zsun, considerably larger than observed systems.  In addition, we find star formation is truncated between redshifts 4 and 7, at odds with the observed late time star formation in isolated dwarf systems but in agreement with Milky Way ultrafaint dwarf spheroidals.  We investigate the efficacy of energetic feedback in our simple thermal-energy driven feedback scheme, and suggest that it may still suffer from excessive radiative losses, despite reaching stellar particle masses of about 100 \Msun, and a comoving spatial resolution of 11 pc.
\end{abstract}


\section{Introduction}

Dwarf galaxies are the most dark matter dominated objects in the universe, with gravitational masses a hundred or even a thousand times greater than their observed baryonic masses.  They therefore appear to lie at the extreme limit of galaxy formation.  Despite this, however, they exhibit remarkable consistency in many of their properties.  Milky Way dwarfs appear to have a constant central mass (within 300 pc) of $10^7$ \Msun\ over five orders of magnitude in luminosity \citep{strigari08} and their mean stellar metallicity tightly correlates with their luminosity over more than three orders of magnitude \citep{kirby11a}.  Despite their small sizes and large mass-to-light ratios, dwarf galaxies appear to lie on many of the scaling relations seen in massive galaxies \citep{kormendy09,tolstoy09}.  Understanding the physics that shape dwarf galaxies is partially an effort to understand a particularly inefficient mode of galaxy formation, but it also is an effort to understand how physical processes involved in galaxy formation scale across galaxy mass to produce the observed continuous evolution in galaxy properties.

The competition between a variety of baryonic heating and cooling mechanisms plays an important role in regulating the gas available for star formation in galaxies.  Understanding galaxy formation is therefore in part about understanding how these microphysical processes work in galaxies.  These processes operate in a non-linearly growing potential well while being perturbed by environmental effects such as ram pressure and by effects of secular evolution of massive baryonic structures such as disks and bars.  Since dwarf galaxies in the Local Group, especially dwarf spheroidal galaxies (dSphs), are observed to have large mass-to-light ratios and old stellar populations, it is likely that their evolution has been heavily influenced by heating processes that suppressed later star formation.  In addition, the number of dwarf galaxies observed around the Milky Way is several orders of magnitude less than the predicted number of dark matter subhaloes found in cosmological N-body simulations \citep{klypin99,moore99}.  This 'missing satellite' discrepancy had been somewhat alleviated by the discovery of many `ultrafaint' dSphs \citep[e.g.][]{willman05} and by hydrodynamical simulations of dwarf satellite systems that have shown baryonic processes can suppress star formation \citep{okamoto10,wadepuhl11}.  Several heating processes are at play in dwarf evolution, including the metagalactic UV background during the epoch of reionization \citep{bullock00, gnedin06}, supernova feedback \citep{governato10,sawala10} as well as local UV heating and winds from young stars. 

While these heating processes act in the evolution of galaxies of all sizes, dwarf galaxies provide an attractive laboratory for their study since their small sizes allow simulators the ability to reach higher resolutions.  High resolution is important for resolving the dense gas from which stars form and for retaining the energy in hot gas, and is therefore key to constraining the star-formation rate.  High spatial resolution (of order 1 pc) is especially important in limiting diffusion from cold, high-density gas into hot, rarefied gas produced by supernova explosions \citep[e.g,][]{deavillez02, joung06}.  Without resolution on this scale, feedback energy is rapidly radiated \citep[e.g.][]{katz92}.  Simulations of galaxy formation often compensate for this behavior by imposing feedback driven winds or by enhancing the impact of feedback heating by either distributing energy to diffuse gas phases \citep[e.g.][]{springel03, scannapieco06} or turning off cooling for a time \citep{stinson06, governato10}.  These different approaches to modeling feedback can produce different star formation histories and galaxy properties even when keeping the total amount of supernova energy the same between different model implementations \citep{schaye10,sales10}.  We understand a great deal about the structure of supernova remnants from observations of nearby resolved examples \citep[e.g.][]{badenes10}, and it is possible to model the impact of supernovae in idealized simulations of the ISM \citep[e.g,][]{deavillez00, joung06}, but how these processes depend on, and interact with, global galaxy properties to produce global behavior such as galactic winds remains an open question.

Heating mechanisms appear to be very important in dwarfs, but understanding the cooling mechanisms that balance them is also crucial.  Many of the heating processes we have outlined are a consequence of stellar evolution and are therefore correlated with the star formation rate.  In low mass haloes with very low metallicities, the effect of molecular line cooling from H$_2$ may be an important factor in determining the star formation rate.  Cosmological galaxy simulations often assume an equilibrium cooling curve for low metallicity gas, but non-equilibrium effects may also be important \citep{abel02, gloverclark12}.  

As we have discussed, reproducing the proper phase distribution of gas in the interstellar medium is quite difficult, however, modeling star formation from this medium through the creation of aggregate star particles is also quite uncertain.  We see stars forming from dense molecular clouds in our own galaxy and \citet{kennicutt98} has demonstrated that the star formation rate of galaxies correlates strongly with the gas surface density.  Despite this observed relation, the gas density threshold for star formation is often an adjustable parameter in cosmological galaxy simulations in part because of resolution limitations, and indeed, adjusting the star formation threshold at fixed resolution can produce very different galaxy properties \citep[see][]{guedes11}.  Modeling star formation from particular gas phases may also produce different behavior; \citet{kuhlen12} have shown that modeling star formation as dependent on the molecular gas fraction suppresses the star formation rate in low mass galaxies at high redshift.  However, \citet{glover12} have argued that simply the gas density determines the star formation rate and the molecular gas fraction plays little direct role.  Since many of the other uncertain subgrid processes that we have discussed (i.e. feedback) are dependent on the star formation rate, uncertainties in modeling star formation have the potential to compound themselves in a galaxy model, and in low mass dwarf galaxies where star formation and its quenching can balance so precariously, these issues may be especially important.

For this study, we have deliberately chosen simple and widely used prescriptions to model subgrid baryonic effects such as supernova feedback and reionization.  Our goal is to test these model prescriptions at high resolution by carefully comparing our simulation results to observations.  We believe that the specific ways in which our simulations succeed or fail to reproduce observations can tell us much about the ability of these model prescriptions to capture the relevant baryonic physics properly.

We will also attempt to better understand the factors that control the gas content, star formation rate, and stellar mass for Local Group dwarf analogues.  We are particularly interested in probing the relative impacts of supernova feedback and reionization on the evolution of these low-mass systems.  We will attempt to do this by better understanding the detailed evolution of the gas phase distribution within a cosmological context.  The simplicity of our model assumptions for subgrid baryonic physics will hopefully provide clarity in interpreting the implications of our results for the evolution of real systems.

To these ends, we have conducted a series of high resolution cosmological, hydrodynamical simulations of a single dwarf halo with prescriptions for H$_2$ and metal line cooling; star formation and supernova feedback; and a global UV background that ionizes neutral gas and dissociates molecules.  We target a dark halo mass of $10^9$ \Msun, at the low-mass end of the range in dwarf galaxy halo masses.  Modeling of the dynamics of Local Group dwarfs suggests they reside in dark matter haloes of approximately this mass or smaller \citep[e.g][]{tollerud11}.

We describe our simulation methods and physical prescriptions in Section \ref{sec:methods} and present the results of our canonical simulations in section \ref{canonical_runs}.  We have also conducted simulations where we have adjusted some of the physical prescriptions implementing supernova feedback and reionization to gauge their effect in our model.  We present the results of these alternate physics runs in Section \ref{alternate_physics}.  We discuss the results of our simulations and compare our results to observations and other simulations in Section \ref{discussion}, and summarize our conclusions in Section \ref{conclusions}.

\section{Methods}
\label{sec:methods}

We conducted our simulations with the structured adaptive mesh refinement (AMR) hydrodynamics code Enzo \citep{bryan99,normanbryan99,o'shea04}.  Enzo uses an Eulerian method for solving the hydrodynamics equations on a Cartesian grid.  We chose a grid-based method because of its ability to resolve interfaces in multiphase gas \citep{slyz05, agertz07,tasker08}.  We used the ZEUS hydro solver in Enzo which uses a simple and robust second-order finite difference method \citep{vanleer77,stonenorman92}.  Dark matter and stars are represented by particles, and the gravitational interactions are computed by solving the potential on the mesh \citep{o'shea04}.  Our model also solves non-equilibrium evolution rate equations for e$^-$, H, H$^+$, He, He$^+$, He$^{++}$, H$^-$, H$_2$, and H$_2^+$, as well as cooling from these species \citep{anninos97,abel97}.  In addition, we include metal line cooling, photoionizing and photodissociating backgrounds, star formation and thermal feedback, and self-shielding by neutral and molecular gas, which are all described in more detail below.  Our version of the code is publicly available in the auto-inits branch of the online Enzo repository\footnote{enzo-project.org} as changeset 15651fe320ff.

\subsection{Initial conditions}

We generated cosmological initial conditions with Enzo's initial conditions generator {\it inits}.  We used the best fit cosmological parameters from the WMAP5 data release ($\Omega_m=0.274$, $\Omega_b=0.0456$, $\Omega_{\Lambda}=0.726$ and $h=0.705$) \citep{hinshaw09}.  We begin our simulations at redshift $z=99$ when perturbations in the universe still grow linearly.  Our cosmological box is 4 comoving Mpc h$^{-1}$ on a side and is divided into $128^3$ cells on the root grid.  We note that our box size is too small to accurately sample large scale modes, and the largest mode in the box becomes non-linear at low redshift.  This will affect the statistics of the haloes in our box, and will also impact accretion at late times; however, since this is an exploratory study to understand the gross evolutionary properties of a single low mass dwarf halo, we argue that this level of error is acceptable.  In particular, we note that while the accretion history of our chosen halo may be biased, it is likely to fall within the wide range of possible such histories for haloes of this mass.  We will redress this problem in later work.
 
Our goal is to simulate a single dwarf halo at high resolution.  We first selected a target halo from a simulation of our cosmological box with no refinement.  We identified haloes close to our target mass of $10^9$ \Msun\ at $z=0$.  We found that lower mass haloes traveled a longer distance relative to their size than higher mass haloes (reflecting the fact that all haloes participate in large-scale bulk flows).  We chose a halo that traveled a relatively small distance relative to its $r_{200}$ radius during the simulation to minimize the area of refinement necessary in resimulations.  Our chosen halo travels a distance of 0.9 comoving Mpc h$^{-1}$ during the simulation.   This choice was motivated by concern for computational efficiency, however, it had the result of selecting a relatively isolated halo.  The closest halo with comparable mass to our target halo has a mass of $8.52 \times 10^8$ \Msun\ and is at a distance of 338 kpc, which is approximately 14 times $r_{200}$ for our halo.  The most massive halo in the box is $1.43 \times 10^{12}$ \Msun\ and is 2.9 Mpc from our target halo. 

The target halo identified, we then created new, refined initial conditions based on the trajectory of our halo and its particles.  We laid down a series of nested static subgrids that refine the root grid by a total of three additional levels so that in the most refined region the effective initial resolution was $1024^3$.  The cell size within the most refined static subgrid before adaptive refinement is therefore 5.54 comoving kpc.  One corner of the maximally refined static subgrid was defined by the target halo's final position plus a buffer of 10 $r_{200}$ radii.  The opposite corner was set to contain the volume enclosing the initial positions of the halo's particles in the coarse simulation, plus an additional buffer of 10\% of the width of this enclosing volume in each direction.  A buffer region of four intermediate resolution grid cells was placed around the boundaries of each successively nested static grid.  The initial resolution of a cell sets the mass of the dark matter particles laid down within it, so the higher resolution cells have lower mass dark matter particles.  The buffer regions are important to ensure that higher mass particles did not impact the evolution inside the highly refined region. These initial conditions set the minimum mass of dark matter particles to be 5353 \Msun.

To further restrict refinement to the halo of interest, we have employed a new technique that identifies dark matter particles in our refined initial conditions that end the simulation in our target halo and then requires the presence of these particles for the refinement of a given cell.  We identified halo dark matter particles in our refined initial conditions by first running a dark matter only simulation to $z=0$.  We identified particles that ended the simulation within 3$r_{200}$ radii of the halo centre and tagged them as `must refine particles.'  In all subsequent runs with hydrodynamics, we restrict adaptive refinement to cells that contain must refine particles and that are within the highest resolution initial subgrid.  Cells meeting these criteria may be further refined based on their dark matter and baryon density such that additional refinement was added (by a factor of 2) whenever the dark matter mass in a cell exceeded four times the dark matter particle mass, with a similar criterion for the gas. We did not explicitly implement refinement based on the Jeans length, however, we found that with our density refinement criteria that in most dense regions we either resolved the Jeans length by four grid cells (meeting the criterion of \citet{truelove97} to avoid artificial fragmentation) or the density of the gas was greater than our chosen threshold density for star formation.  The net result of our method is that the adaptive refinement tracks the dark matter particles and gas that ends up in the halo of interest.  In our highest resolution run (Simulation R10 in Table \ref{tab:summary}), we find that 99\% of the dark matter particles within $r_{200}$ are must refine particles at $z=0$.  This method resulted in a computational savings of more than a factor of 10, as compared to allowing refinement anywhere in the most refined initial grid, which is standard practice.

In our highest resolution simulations, we allow a total of 12 levels of adaptive refinement beyond the root grid.  These simulations therefore have a minimum cell length of 10.8 comoving pc.  This means at high redshift ($z\sim9$) our calculations have a physical resolution of order one pc.  We have also conducted simulation R43 that allows only 10 levels of adaptive refinement beyond the root grid for a minimum cell length of 43.3 comoving pc.  We will discuss this simulation in Section \ref{sec:resolution}.

\subsection{Metal cooling}

As noted earlier, we include radiative cooling from nine atomic and molecular species \citep{abel97, anninos97}, whose (non-equilibrium) abundances are explicitly followed.  In addition, the code also tracks a mean metallicity, and additional metal heating and cooling rates are interpolated from equilibrium cooling tables, in a way similar to that described in \citet{smith08}: Cloudy \citep{ferland98} is used to generate five-dimensional tables that give the net metal cooling and heating rates, depending on the gas metallicity, electron fraction, density, temperature and redshift.  The redshift tracks the time-dependence of the metagalactic UV background (see below). The combination of heating and cooling from non-equilibrium atomic and molecular hydrogen combined with equilibrium metal lines, allows for a more accurate estimate of the cooling rates, particularly for low metallicity gas.  Since we follow the abundance of H$_2$, the cooling rates we compute for low temperature, low metallicity gas where H$_2$ cooling dominates are particularly accurate.  We do not include molecular hydrogen formation on dust grains, but by the time this pathway becomes important, metal cooling dominates over H$_2$ cooling anyway \citep{glover03}.  Finally, we note that we do not include molecules formed out of heavy elements, but these tend to be less important than fine-structure lines \citep{gloverclark12}, particularly at low abundances.

\begin{table*}

 \begin{minipage}{100mm}
  \caption{Summary of Simulations. \label{tab:summary}}
  \begin{tabular}{@{}cccccc@{}}
  \hline
   Name & $\Delta$x$_{min}$ & $e_{SN}$ & $\Delta z$ & H$_2$  & $Z_{\rm{max}}$\\
    & (comoving pc) & & & Cooling  & (\Zsun)\\
  \hline
	R10              & 10.8 & $3.7\times10^{-6}$ & 7-6   & yes & none 	\\
	R10-earlyUV      & 10.8 & $3.7\times10^{-6}$ & 9-8.9 & yes & none 	\\

	R10-noH2         & 10.8	& $3.7\times10^{-6}$ & 7-6   & no  & none	\\ 

	R10-noUV         & 10.8	& $3.7\times10^{-6}$ & none  & yes & none	\\
	
	R10-noFB	 & 10.8	& $0$		    & 7-6   & yes & none	\\
	R10-noFB-LimCool & 10.8	& $0$		    & 7-6   & yes & 0.1		\\
	R10-lowFB	 & 10.8	& $10^{-6}$	    & 7-6   & yes & none	\\

        R10-DM           & 10.8 & -                 &  -    &  -  & - 	        \\
	
	R43	         & 43.3	& $3.7\times10^{-6}$ & 7-6   & yes & none	\\
 \hline
\end{tabular}
Note: The quantities presented in each column are (1) the minimum cell width for the calculation, (2) the fraction of rest mass energy that star particles return to the gas, (3) the redshift range over which the UV background is introduced, (4) whether the abundance of  or cooling from H$_2$ is tracked, (5) and the maximum effective metallicity for metal cooling.
\end{minipage}
\end{table*}

\subsection{UV background}

We include both photoionizing and photodissociating metagalactic backgrounds in our model.  We assume spatially uniform photoionizing rates for H I, He I and He II that vary with redshift, taken from \citet{haardtmadau01}.  This background includes ionization from both quasars and stars.  For most of our runs we start turning on the photoionizing background at $z=7$ and gradually ramp it up to full strength by $z=6$.  The photodissociation rate of H$_2$ is taken to be $1.13\times10^{-8}$ times the flux in the Lyman-Werner bands \citep{abel97}.  We adopt the average flux in the Lyman-Werner bands from the spectra of \citet{haardtmadau11}.  This flux is redshift dependent and we assume it to be spatially uniform.  We turn on the photodissociating background at the same time as we start the photoionizing background (without a ramp-up period, as there is no significant gas heating associated with this transition).  In order to investigate the impact of the timing of the reionization epoch, we also carried out a simulation in which we turned on the UV background at an earlier redshift, $z = 8.9$, and ramped up the ionizing background to full strength by redshift $z=8$.

Although the UV background is spatially uniform, we apply a crude self-shielding correction.  The H I ionizing flux felt by each cell is attenuated by a factor of $\exp(- \Delta x n_{\rm HI} \bar{\sigma}_{\rm ion})$, where $\Delta x$ is the cell size, $n_{\rm HI}$ is the neutral hydrogen density, and $\bar{\sigma}_{\rm ion}$ is the frequency-averaged photoionization cross-section for H I.  We computed the frequency-averaged photoionization cross-section as

\begin{equation}
\label{eq:photoionization-cross-section}
\bar{\sigma}_{\rm ion} = \frac{\int\limits_{\nu_0}^\infty \frac{4\pi J}{h\nu}\sigma\,\mathrm{d}\nu}{\int\limits_{\nu_0}^\infty \frac{4\pi J}{h\nu}\,\mathrm{d}\nu},
\end{equation}

\noindent
where $J$ is the flux of the ionization background, which we approximate as $J \propto \nu^\beta$, and $\sigma$ is the frequency dependent ionization cross-section, which is zero for frequencies less than $\nu_0$, the threshold frequency for ionization of H I, and $\sigma \propto A_0 \nu^{-3}$ for frequencies greater than $\nu_0$.  The parameter $\beta$ is the power law slope of the ionization background which we take to be -1.57, approximately the slope of the photoionization background that we use.  The parameter $A_0$ is the peak ionization cross-section for H I.  With these assumptions we calculate the average photoionization cross-section to be $\bar{\sigma}_{\rm HI} = A_o \beta / (\beta - 3)$.

A similar correction is made for the H I photoheating rate, where we use the frequency-averaged photoheating cross-section which we compute as

\begin{equation}
\label{eq:photoheating-cross-section}
\bar{\sigma}_{\rm heat} = \frac{\int\limits_{\nu_0}^\infty \frac{4\pi J}{h\nu}(h\nu - h\nu_0)\sigma\,\mathrm{d}\nu}{\int\limits_{\nu_0}^\infty \frac{4\pi J}{h\nu}(h\nu - h\nu_0)\,\mathrm{d}\nu},
\end{equation}

\noindent
which we find is proportional to $\bar{\sigma}_{\rm ion}$ by a factor of $(\beta+1)/(\beta-2)$.  Corrections are also made for He I and He II ionization and heating using the corresponding values of $A_0$ for each species taken from \citet{brown71} and \citet{osterbrock}.  These corrections are approximate, as they assume that only gas in the local cell contributes to self-shielding, but they do suppress ionization and heating in dense regions.

We also apply a simple self-shielding correction to the Lyman-Werner background as described in \citet{shang10}; this uses the local H$_2$ densities multiplied by the local Jeans length to determine an estimate of the column density of H$_2$ that is shielding each cell.  The simple estimate in \citet{draine96} is then used to compute a self-shielding correction factor.  \citet{wolcottgreen11} have recently shown that this can result in an overestimate of the self-shielding correction for our conditions; however, given the highly uncertain level of the Lyman-Werner background, this approximation is sufficient for this work.

\subsection{Star formation and supernova feedback}

Star particles are created during the course of a simulation following the prescription of \citet{cenostriker92}.  The implementation in Enzo is described in more detail in \citet{tasker06}, but we briefly outline it here.  Star particles are created in a cell if its density rises above $10^5$ times the universal mean density at that redshift, if the divergence of the local velocity is negative, and if the local cooling time is less than the dynamical time (or the gas temperature is less than $10^4$ K).   If these requirements are met, then the star formation rate is computed according to $\dot{\rho}_{\rm SFR}  = \epsilon_{\rm SF} \rho_{\rm gas} / t_{\rm dyn}$, where $\epsilon_{\rm SF} = 0.01$ is the star formation efficiency (e.g. \citet{krumholztan07}) and $t_{\rm dyn} = (3 \pi / 32 G \rho)^{1/2}$ is the dynamical time.  A star particle is then created with mass $M_* = \dot{\rho}_{\rm SFR} \Delta t \Delta x^3$, where $\Delta t$ is the timestep.   To prevent very small star particle masses, a stochastic prescription is used such that star particles are not created until 100 \Msun\ of stars are predicted to form; once that limit is reached, a star particle of 100 \Msun\ is created (up to a maximum of 80\% of the baryonic mass in the cell).

Once they are created, star particles return both energy and mass to their surrounding gas, also following the prescription of \citet{cenostriker92}.  Star particles return some fraction of their rest mass energy to thermal energy in the gas.  We chose this value to be $3.7\times10^{-6}$ for most of our simulations.  This corresponds to 150 \Msun\ worth of stars to produce one supernova's worth of energy ($10^{51}$ ergs).  This energy is continuously injected as thermal energy into the gas over the feedback time $t_{f}$, which is a few dynamical times (i.e. ten Myr to a few tens of Myr) as detailed in \citet{tasker06}.  The energy and mass return fractions depend sensitively on assumptions regarding the IMF; the values adopted in this paper are in the ranges determined by other works \citep[e.g.][]{jungwiert01,kroupa07,oppenheimer08,conroy10}.  Note that we do vary the fraction of rest mass energy returned in feedback in order to investigate the impact this has on our results.  The particle also returns 25\% of its mass to the gas.

Our choice of a purely thermal feedback model is motivated by simple models of individual supernovae which show that the supernova driven Sedov blast wave reaches a radius of a few parsecs in dense gas before radiative cooling begins to slow it down and it radiates away energy \citep{chevalier74}. One might expect that in a galaxy simulation with a resolution comparable to this scale, artificial radiative losses seen in previous lower resolution studies \citep[e.g.][]{katz92} would be small. In such a simulation this resolution would negate the need for artificial subgrid models necessary to compensate for this effect (e.g. turning off radiative cooling). Our main simulations have a spatial resolution of 10.8 comoving pc; therefore, for $z \ge 3$, our simulations have a spatial resolution less than 3 pc, which is the level of resolution needed to test this hypothesis. Given the observed old stellar populations found many dSphs and their small sizes, these systems are an ideal test case for this type of model run at these resolutions.  We will discuss the efficacy of this model and our assumptions in Section  \ref{sec:feedback_model}.

We follow a single gas metallicity field in all our simulations, so we do not follow the detailed chemical evolution of individual metal species.  The metallicity field ($\rho_Z$) is initialized to a low value at the beginning of the simulation ($10^{-10}$ times the mean gas density).  When star particles form, they are assigned the metal fraction $\rho_Z/\rho_{gas}$ of their birth cell.  Stars return 2\% of their total mass to this metallicity field over the feedback timescale $t_{f}$ using the same temporal functional form as used for the thermal feedback.  During each time step, the ejected metal mass is added directly to the single cell in which the star particle resides. The 2\% metal yield we use is a fiducial value consistent with previous work \citep{madau96}.  Uncertainties in individual stellar yields and the IMF make this value very difficult to constrain.  Moreover, a more sophisticated chemical model could be employed that includes different types of supernovae with different injection timescales and chemical yields.  We are interested in probing the broad effect of a using a purely thermal supernova model at high resolution in an AMR code and understanding the \textit{relative} effects of supernovae and reionization.  For these reasons we have chosen to use a simple chemical model.

\subsection{Halo tracker and data analysis}

The problem of finding and tracking inheritance between dark matter haloes within cosmological simulations is not a trivial one.  We used the halo finder $\tt{hop}$ \citep{eisenstein98} to identify haloes in our simulation and a simple halo tracker to track inheritance between haloes, starting with the final target halo at $z=0$ and working backwards in time to higher redshifts.  For each child halo, the tracker finds parent haloes in the preceding output by identifying haloes that contain dark matter particles that end up in the child halo.  Parent haloes that contribute more than 10\% of the child halo's dark matter mass or contain more than $10^7$ \Msun\ in dark matter mass are considered significant parents.  The lineages of these significant parents are then tracked back to previous outputs in the same manner.

The ability of this simple halo tracker to separate halo lineages is limited by the ability of $\tt{hop}$ to separate interacting haloes and by the cadence of simulation outputs.  There were cases of interacting haloes where $\tt{hop}$ was unable to properly separate haloes and the halo tracker was therefore unable to properly separate lineages containing these haloes.  In these cases, we present the potential lineages of interacting haloes as equivalent and do not attempt to distinguish between them.

We analysed the baryon quantities of identified haloes with the simulation data analysis and visualization tool yt \citep{turk11}.  We also used yt to construct the images, phase diagrams and radial halo profiles which we present.

\section{Results}
\label{MainResults}

\begin{figure}
\includegraphics[width=84mm]{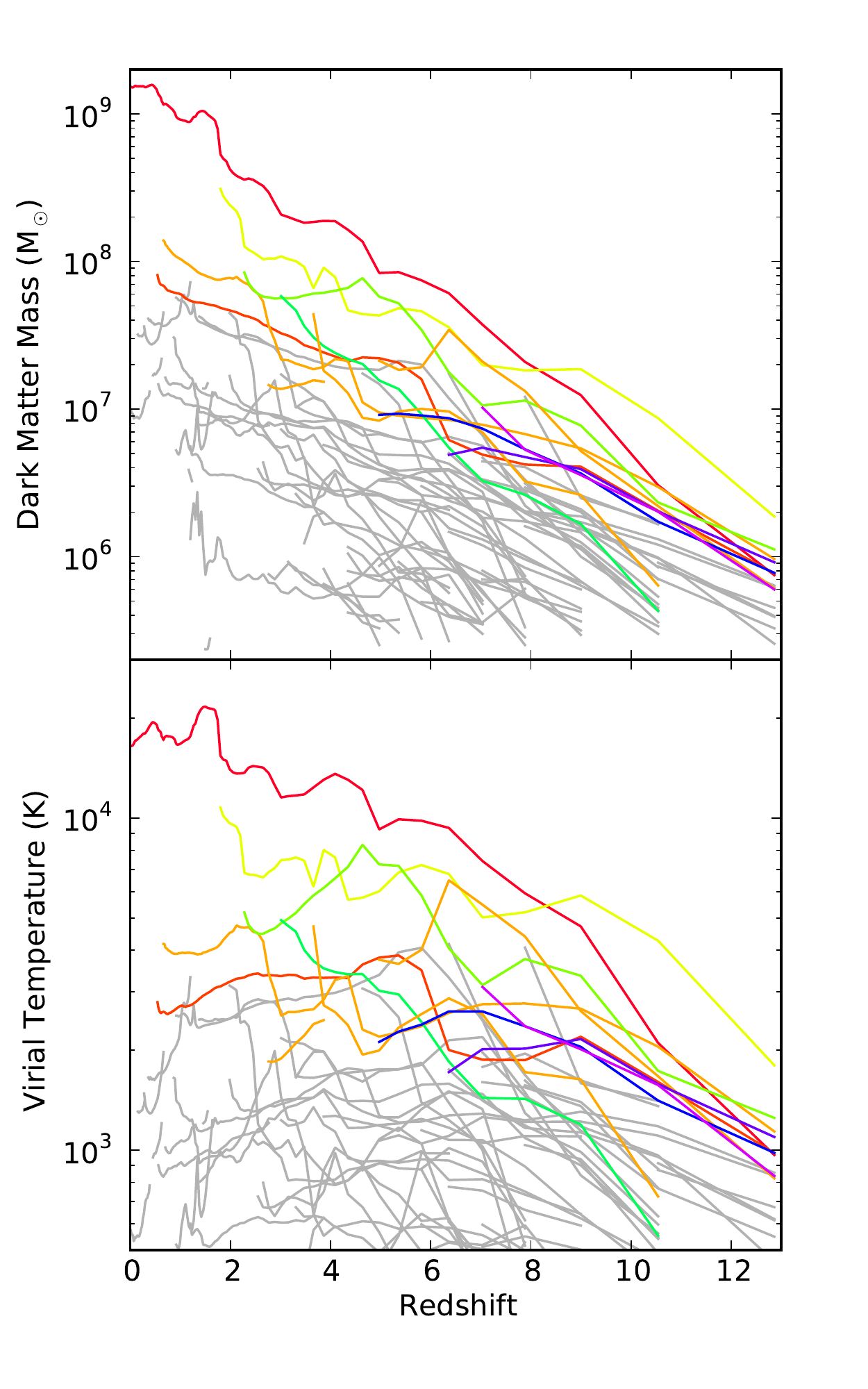}
\caption{Top: Evolution of the dark matter mass within $r_{200}$ (top) and virial temperature (bottom) of progenitor haloes in simulation R10.  Coloured lines indicate haloes containing star particles, while gray lines indicate haloes that remain dark.  A line begins when a halo becomes massive enough to be detected by our halo finder and ends when the halo merges into a more massive halo.  We note that substructure can occasionally separate far enough from its parent halo to be detected for a brief time as a separate halo, and therefore appears as short lines.  In particular, two star forming progenitors have a series of close encounters during which our halo finder was unable to distinguish between them -- their evolution is shown in orange.}
\label{fig:halo_properties}
\end{figure}

We have run a suite of simulations of a single dark matter halo, varying the physical prescriptions outlined in Section~\ref{sec:methods}.  Table~\ref{tab:summary} summarizes the simulations discussed in this paper.  Analysis is conducted on simulation outputs that are written every 108 Myr.  Two of our runs, R10 and R10-earlyUV, include all of the physics we have described previously, but differ in the time at which the UV backgrounds are introduced.  We call these two runs our canonical runs, and describe them in detail below.  We have also run a simulation completely neglecting the UV backgrounds, simulation R10-noUV.  This simulation, unlike our other simulations which were run to $z=0$, was run only to $z = 3.1$.  We chose to end R10-noUV at this redshift because of the copious cold, dense gas produced.  This gas made the simulation run quite slowly and by redshift 3.1, the difference in evolution was clear.  Simulations R10-lowFB, R10-noFB and R10-noFB-LimCool all examine the impact of changing the strength of supernova feedback.  We also tested the importance of H$_2$ for cooling by removing it as a coolant in simulation R10-noH2.  This set of simulations gives us broad levers on the relative importance of most of the heating and cooling effects important in dwarf formation.  In particular, we aim to understand the fundamental question of how the dual effects of supernova feedback and reionization shape the baryon fractions and star formation rates of dwarf galaxies.

\subsection{Canonical runs}
\label{canonical_runs}

\begin{figure*}
\centering
\includegraphics[width=160mm]{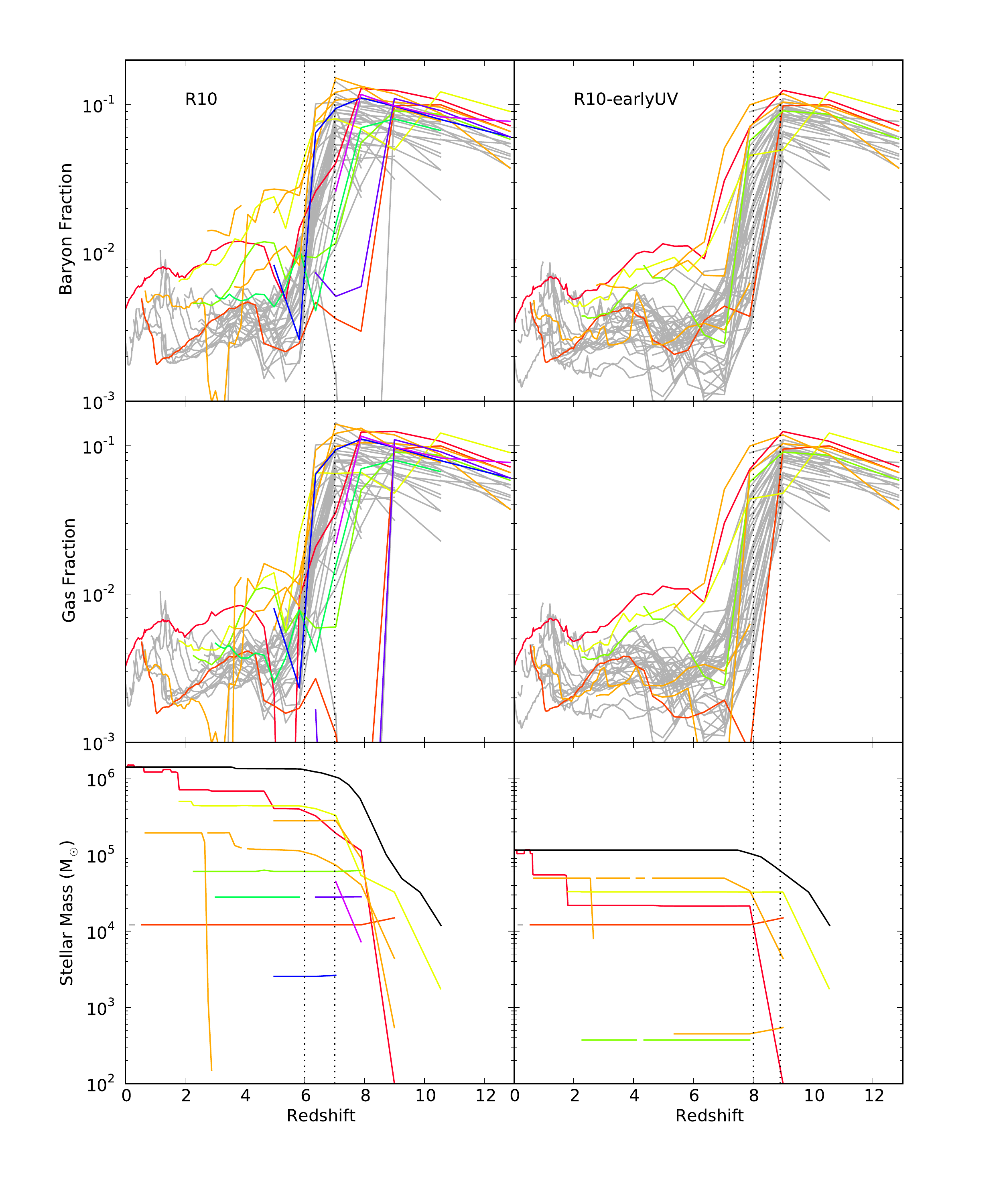}
\caption{Evolution of baryon properties, within $r_{200}$, of progenitor haloes in simulation R10 (left column) and R10-earlyUV (right column).    Top row: The ratio of baryon mass to total mass.  Middle row: The ratio of gas mass to total mass.  Bottom row: The total stellar mass.  The cumulative mass distribution of star particles that end the simulation in the final halo is shown in black. The solid line colours have the same meaning as in Figure \ref{fig:halo_properties}.  The dotted lines bracket the redshift range over which the photoionization background is introduced and ramped up to full strength.}
\label{fig:baryon_properties}
\end{figure*}

In this section, we describe in detail the results of R10 and R10-earlyUV, which are the highest resolution simulations to include all of the physics outlined in Section~\ref{sec:methods}.  R10-earlyUV differs from R10 in that the uniform UV backgrounds were turned on in the same way as described in Section~\ref{sec:methods} but between redshifts 8 and 8.9 instead of between redshifts 6 and 7.  The purpose of introducing the global UV background at different times is to explore the effect of patchy reionization.  More isolated regions of the universe farther from major sources of ionizing photons may be affected by the ionizing background later than less isolated regions.  In both R10 and R10-earlyUV, the universe is ionized by redshift six.

\subsubsection{Global properties}

The halo we have chosen is fairly isolated at $z=0$; however, like all dark matter haloes in cosmological simulations, it assembles hierarchically.  Figure~\ref{fig:halo_properties} shows the evolution of the dark matter halo masses and virial temperatures \citep[as defined in][]{machacek01} of progenitor haloes in R10 and R10-earlyUV.  We present the dark matter evolution only for haloes in R10 since the dark matter evolution is virtually identical in R10-earlyUV.  At $z=9$, there are 30 progenitor haloes more massive than $10^6$ \Msun.  Two of these haloes are more massive than $10^7$ \Msun, and these two haloes gradually build up their mass within two groups over the course of the simulation.  The two haloes merge at $z= 1.8$ in a merger that is about 2:1 in dark matter and nearly 1:1 in stellar mass in both R10 and R10-earlyUV.  

\begin{figure*}
\includegraphics[width=160mm]{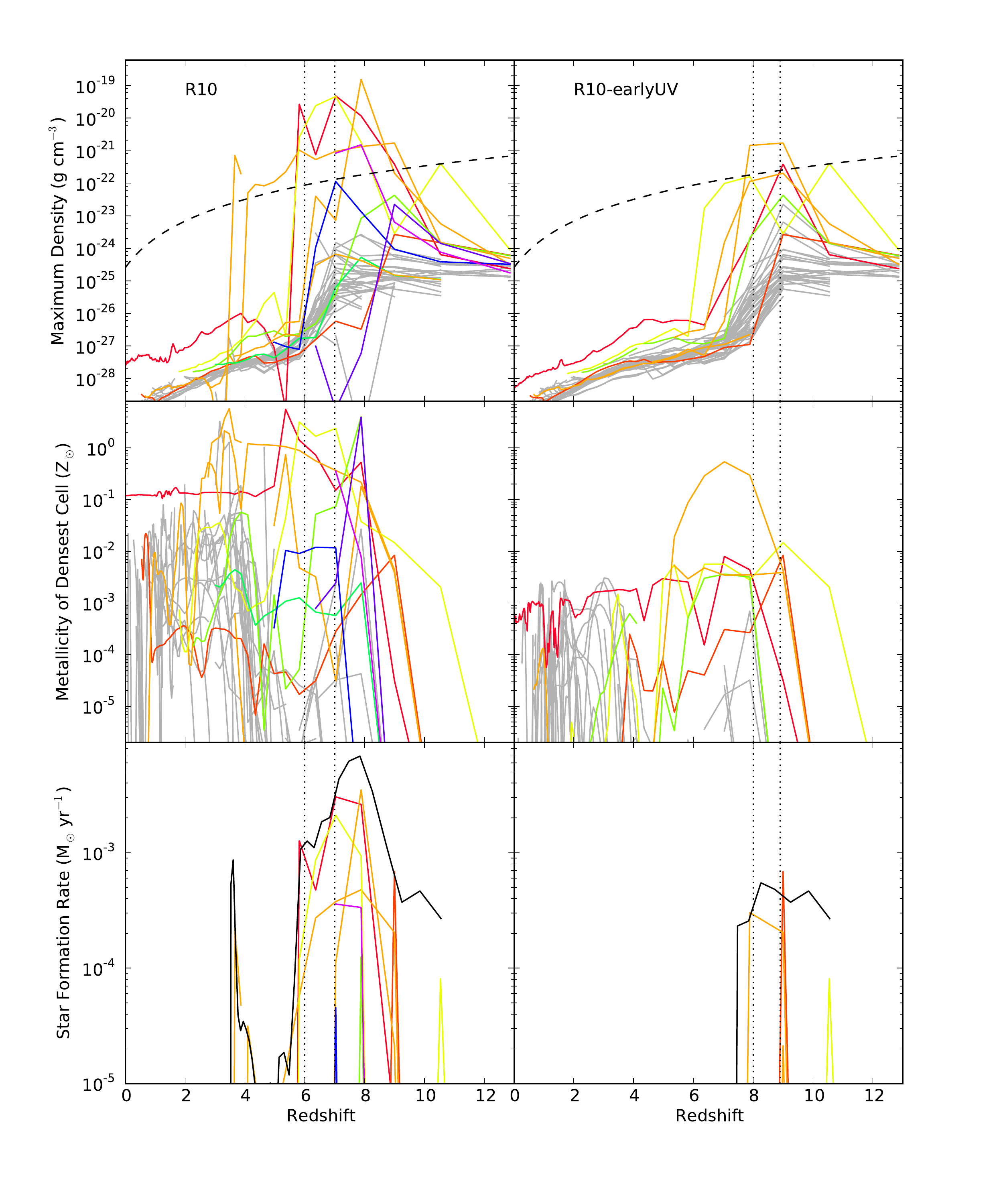}
\caption{Evolution of the peak density, metallicity and star-formation rate within progenitor haloes in R10 (left column) and R10-earlyUV (right column).  Top row: Evolution of the maximum density within each progenitor halo.  The density threshold for star formation is shown as a dashed line.  Middle row: Evolution of the metallicity of the peak density cell shown in the top row.  Bottom row:  The star-formation rate within the last 22 Myr for each progenitor halo is plotted with colour lines, while the star-formation rate inferred from the star particles that end the simulation in the final halo is shown in black.   One halo in R10 (plotted in dark green) underwent a single star formation event that happens between outputs.  Another halo in R10 (plotted in dark purple) accreted its stellar population through a major merger with a slightly less massive progenitor but did not form its own {\it in situ} stellar population before the merger.  Neither halo is therefore plotted in the bottom panel.  A different halo in R10-earlyUV (plotted in light green) also underwent a star formation event between outputs and is not plotted in the bottom panel.}  The dotted lines bracket the redshift range over which the photoionization background is introduced and ramped up to full strength.
  \label{fig:peak_density_cell_properties}
\end{figure*}

The evolution of a variety of baryon quantities in progenitor haloes is shown in Figures~\ref{fig:baryon_properties} and \ref{fig:peak_density_cell_properties}.  We track gross properties of progenitor haloes in Figure~\ref{fig:baryon_properties} such as the total baryon fraction, gas fraction and stellar mass and  quantities associated with the densest cell in each halo in Figure~\ref{fig:peak_density_cell_properties} such as its density and metallicity.  The densest cell within each halo is the cell most likely to form star particles in our model.  Given the low star formation rates expected of progenitor haloes and the fact that our star formation and supernova feedback models operate on a cell-by-cell basis, the properties of the cell with the maximum likelihood of star formation can be very informative.

The evolution of the gas fraction in progenitor haloes appears to be dominated by reionization.  Figure~\ref{fig:baryon_properties} shows sharp declines in the gas and baryon fractions in both R10 and R10-earlyUV at their respective times of reionization.  These declines are due to photo-evaporative outflows triggered by reionization \citep{barkana99,gnedin06}.  We see that once the gas fraction declines during reionization, it remains suppressed for the remainder of the simulation.  We see no evidence for re-accretion of gas once the main halo has been assembled at $z=1.8$.

There are also smaller, but still significant, declines in the gas fraction prior to reionization in several luminous progenitors (Figure~\ref{fig:baryon_properties}).  These declines appear to be correlated with peaks in the star formation rate as shown in the bottom panels of Figure~\ref{fig:peak_density_cell_properties}.  Therefore, we conclude that they are primarily due to supernova feedback.

The gas fraction indicates how the bulk of the gas responds, but star formation depends on the presence of dense gas.  The evolution of the density of the densest cell in each progenitor halo shares some characteristics with the evolution of the gas fraction, but it is not identical.   The peak cell density declines after reionization, but the decline is delayed relative to the declines seen in gas fraction.  This is due, in part, to self-shielding from the UV background.  In R10 particularly, there appears to be an extended epoch of elevated gas density in one progenitor that does not end until $z=4$ and another progenitor is able to briefly re-cool some gas, although this dissipates by $z=3.5$.  We see that the evolution of the peak cell density tracks the star formation rate in individual haloes, also shown in Figure \ref{fig:peak_density_cell_properties}, which is unsurprising as the presence of dense gas controls star formation in our model.

\begin{figure*}
\centering
\includegraphics[width=175mm]{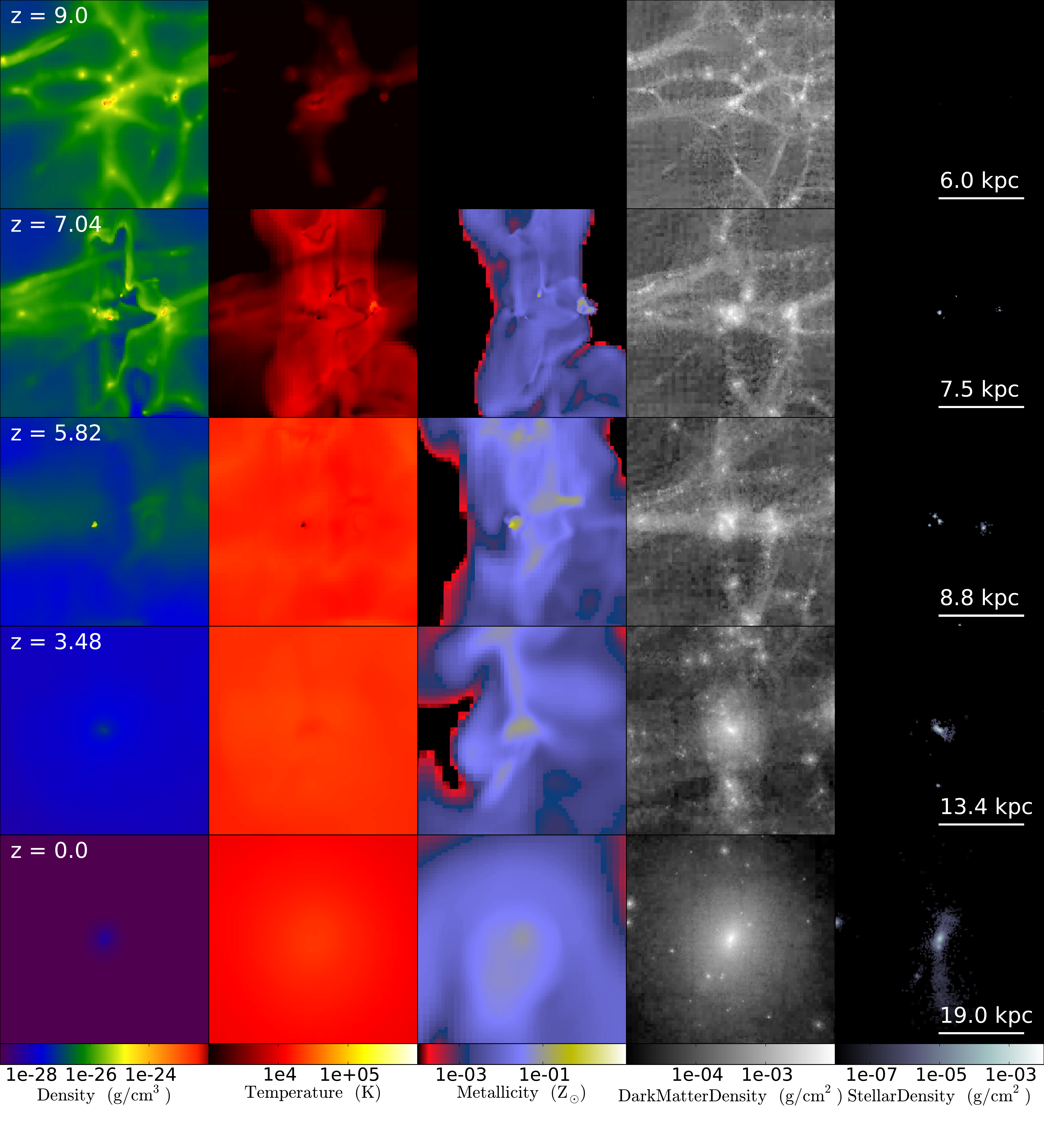}
\caption{From top to bottom, cubic projections centered on a massive star-forming progenitor halo at $z=9.0, 7.04, 5.82, 3.48$ that also show several other lower mass star-forming progenitors.  The bottom row shows the final halo at $z=0$.  From left to right, the panels show projected gas density, temperature, metallicity, dark matter density, and stellar density.  The panels presented in the first four rows have a comoving width of 150 kpc; the final row of projections at $z=0$ have a width of 48 kpc, which is $2r_{200}$ for the final halo.  The scale of the images in physical, non-comoving units is indicated for each redshift.  At $z=9$ the other massive star-forming progenitor is 14 physical kpc from the centered halo and is therefore not shown.  Projections of gas density, temperature and metallicity are weighted by gas density and projections of dark matter density and stellar mass density are unweighted (see Equation 3 in \citet{turk11}).  Note that this weighting scheme gives units of volume density for gas density and units of surface density for dark matter and stellar density.}
\label{fig:image_panel_plot}
\end{figure*}

\begin{figure*}
\centering
\includegraphics[width=100mm]{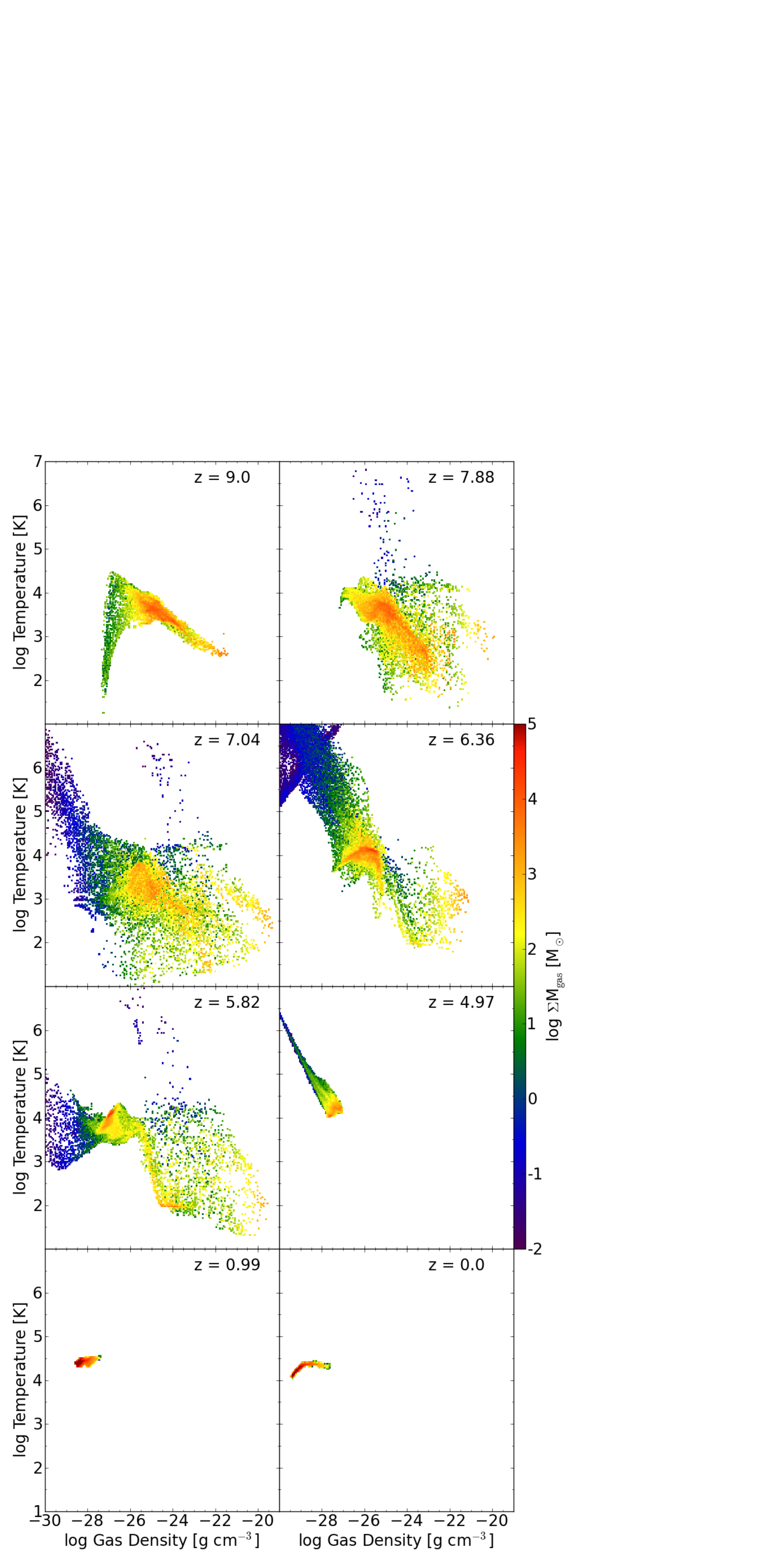}
\caption{Density-temperature distributions of gas within the r$_{200}$ radius for one of the massive halo progenitors at the redshifts shown.  The colour of cells shows the cumulative sum of mass within the corresponding density-temperature bin.}
\label{fig:phaseplots}
\end{figure*}

The evolution of the properties of the densest cell shown in Figure~\ref{fig:peak_density_cell_properties} can be quite stochastic in nature, with sharp changes especially at high redshift.  These changes are due to the short timescale for cell-by-cell variations in our models.  Our high spatial resolution enhances the importance of mixing to the evolution of the ISM, which likely plays a role in these variations.  Also at this epoch, star-formation with accompanying supernova feedback is very important.  At high redshift, regions of high gas metallicity coincide with regions of recent star formation.  Over time these regions of enhanced gas metallicity expand, driven by the energy of supernovae to pollute the IGM as seen in Figure \ref{fig:image_panel_plot}.  This expansion causes the gas metallicity in the central regions where the stars are located to decline.  We discuss the effectiveness of metal ejection in Section \ref{sec:metal_ejection}.

There are multiple star forming progenitor haloes in both simulations; however, R10-earlyUV has fewer, due to the earlier onset of reionization.  It appears that some haloes which cooled later in R10 do not have time to cool before reionization occurs in R10-earlyUV.  The bottom panels of Figure~\ref{fig:peak_density_cell_properties} show both the star formation rate inferred from the stellar population in our target halo at redshift zero, and the instantaneous star formation rates in all progenitor haloes measured at each data output.  The star-formation rate inferred from the stellar population at redshift zero is built up from star-formation events in multiple progenitor haloes with varying durations and intensities. 

Several star forming progenitors undergo star formation events that appear to be single bursts and have a duration of order of or less than the time between data outputs (108 Myrs).  Three star forming progenitors in R10 and four star forming progenitors in R10-earlyUV are captured in a star-forming state in only a single data output.  One progenitor in each of R10 and R10-earlyUV undergoes a star burst that occurs between outputs.  The total mass in stars produced in this mode is small in R10, constituting less than 8\% of the final stellar mass.  In R10-earlyUV its contribution to the final (smaller) stellar mass is much greater, constituting over 57\%.  In R10-earlyUV there is essentially a single progenitor that has an extended period of star formation.

\subsubsection{A detailed look at the evolutionary stages}
\label{sec:detailed_evo}
 
In order to get a better qualitative understanding of how our simulated dwarf galaxy evolves, we next look at the spatial properties of the halo at various times in R10.  The evolution of gross halo properties presented in Figures \ref{fig:halo_properties}-\ref{fig:peak_density_cell_properties} can be tracked through the progression of projections presented in Figure~\ref{fig:image_panel_plot}, which are centered on one of the two massive, star-forming progenitor haloes of our final halo and show its assembly at high redshift from a small group of other dwarf progenitors.  We can see the spatial variations in gas density, temperature and metallicity produced by the processes of cooling, supernova feedback, reionization and self-shielding.  We can also see the hierarchal build-up of both dark matter mass and stellar mass which constitute the final halo.  We can break down the evolution of this halo into a number of states described by these projections and by the density-temperature phase plots for the same progenitor halo presented in Figure~\ref{fig:phaseplots}.
 
The first stage presented is at $z=9$ and is shown in the top left panel of Figure~\ref{fig:phaseplots}.  We see the halo just prior to the onset of star formation.  Near the virial radius the gas shock heats and creates a reservoir of gas at roughly the virial temperature of the halo.  This gas can then begin to cool via H$_2$ cooling \citep[e.g.][]{abel02}.  Figure \ref{fig:halo_properties} shows the evolution of the virial temperature of many progenitor haloes; prior to reionization in both the R10 and R10-earlyUV runs, all progenitor haloes have virial temperatures less than $10^4$ K.  Therefore, for primordial gas in these haloes, the only important coolant is molecular hydrogen; atomic line cooling is not significant.  This stage is also shown in projections in the first row of Figure \ref{fig:image_panel_plot}.  

Eventually the gas reaches the second stage (top-right panel of Figure~\ref{fig:phaseplots}; $z=7.88$) where gas temperatures in the centre fall to 200 K and gas densities cross $10^{-21}$ g cm$^{-3}$, roughly the gas density threshold for star formation in our model at high redshift.  Once gas becomes dense enough to form stars in this second stage, we see evidence of supernova-heated gas with temperatures in excess of $10^6$ K and the picture becomes more complex.

In the third stage (see panel labelled $z=7.04$ in Figure~\ref{fig:phaseplots}), gas continues to be heated by supernovae and shows evidence of pressure equilibrium at low densities and high temperatures, as shown in the phase diagram.  Also, with the injection of metals, gas can now cool to temperatures below 100 K at a broad range of densities.  In some cases, the gas does appear to be significantly affected by supernova feedback in this stage.  Figures \ref{fig:baryon_properties} and \ref{fig:peak_density_cell_properties} show the evolution of the gas fraction and the peak gas density of many haloes.  Prior to reionization there are examples of decrements in both quantities, which must be due to supernova feedback.  The second row in Figure~\ref{fig:image_panel_plot} also shows this stage.  We can see that, by this point, metals have been ejected from progenitor haloes and have significantly enriched the intra-group medium.  

After $z=7$ (in the R10 run), the onset of photoionizing and photodissociating backgrounds causes a dramatic change in the phase structure of gas within progenitor haloes.  In this fourth stage (see panels labelled $z=6.36$ and $z=5.82$ in Figure~\ref{fig:phaseplots}), we see nearly all of the gas heated to a temperature of about $10^4$ K.  In some haloes (including the one shown in Figures \ref{fig:image_panel_plot} and \ref{fig:phaseplots}), the gas is able to remain dense or briefly recool, producing an additional wave of star formation.  The third row of Figure~\ref{fig:image_panel_plot} shows this stage.  This stage does not appear in haloes in R10-earlyUV and it ends in all progenitor haloes in R10 by $z=3.5$. 

After this point, once the dense star-forming gas has been dissipated, a small amount of high-temperature gas persists within the supernova heated phase in a fifth stage (see panel labelled $z=4.97$ in Figure~\ref{fig:phaseplots}).  However, this gas phase eventually dissipates and the small amount of remaining gas sits at a temperature slightly above $10^4$ K.  This is both the temperature of the IGM and roughly the virial temperature of the final halo.

At low redshift (see the last two panels of Figure~\ref{fig:phaseplots}), a series of dry mergers continue until the final halo is assembled.  The final haloes that result in R10 and R10-earlyUV are gas poor spheroids of dark matter and star particles.  R10 and R10-earlyUV differ in stellar mass by an order of magnitude: the halo in R10, which had a late onset of reionization, has a final stellar mass of $1.43\times10^6$ \Msun, while the halo in R10-earlyUV, which had an earlier onset of reionization, has a final stellar mass of $1.16\times10^5$ \Msun.  Table \ref{tab:final_properties} summarizes some of the final halo properties in these simulations.

\begin{table}

\centering
  \caption{Summary of Final Halo Properties. \label{tab:final_properties}}
  \begin{tabular}{@{}cccc}
  \hline
   & R10 & R10-earlyUV  \\
  \hline
$M_{tot}/M_\odot$ & $1.55 \times 10^9$ & $1.55 \times 10^9$ \\
$M_*/M_\odot$ & $1.43 \times 10^6$ & $1.16 \times 10^5$ \\
$M_{gas}/M_{tot}$ & $3.24 \times 10^{-3}$ &  $3.25 \times 10^{-3}$\\
$r_{200}$ (kpc) & 23.7 & 23.9 \\
$r_{1/2}$ (pc)&704 & 213 \\
$M_{1/2}/M_\odot$ & $3.05 \times 10^7$& $3.86 \times 10^6$ \\
$M_{300}/M_\odot$ &$7.53 \times 10^6$& $7.41 \times 10^6$ \\

$\sigma_{1/2}$ (km/s) & 7.83 & 8.30  \\

Log($\langle Z/Z_\odot \rangle$) (median)& -0.29 & -1.22 \\
Log($\langle Z/Z_\odot \rangle$) (mean)& -0.076 & -0.92 \\
$\sigma_Z/Z_\odot$& 0.84& 0.14 \\

 \hline
\end{tabular}

Note: The quantities presented in each row are (1) the total mass within $r_{200}$, (2) the total stellar mass within $r_{200}$, (3) the total gas fraction within $r_{200}$, (3) $r_{200}$, the radius within which the mean halo density is 200 times the critical density of the universe, (4) the radius enclosing half the stellar mass, (5) the total mass within $r_{1/2}$, (6) the total mass within 300 pc, (7) the velocity dispersion of star particles within $r_{1/2}$, (8) the mass-weighted median of the star particle metallicities, (9) the mass-weighted mean of the star particle metallicities, (10) the mass-weighted standard deviation of the star particle metallicities.

\end{table}

\subsection{Alternate physics runs}
\label{alternate_physics}

We have also conducted a number of simulations in which we have altered some aspect of our physical prescriptions in order to better understand their effect on the evolution of the final halo.  We have done simulations that have changed some aspect of the cooling prescription, the supernova feedback prescription and the UV background prescription.  Table \ref{tab:summary} summarizes these simulations.

\subsubsection{Simulation without H$_2$ cooling}

We conducted a simulation without the creation of or cooling from H$_2$.  Simulation R10-noH2 has line cooling from six atomic species of hydrogen and helium as well as metal line cooling and heating from the Cloudy models of \citet{smith08}.  The UV backgrounds, the self-shielding by neutral gas, and the star formation and feedback models are the same as R10.  This simulation produced an entirely dark halo; no star formation occurred.  We found that without molecular hydrogen cooling, gas temperatures in the most massive progenitors did not fall below 500 K and densities remained below $6\times10^{-24}$ g cm$^{-3}$.  The peak in gas density occurred at $z=5.8$, shortly after reionization reached full strength.  Our model requires gas to be denser than $8.09\times 10^{-23}$ g cm$^{-3}$ to form star particles at this redshift.  With no initial star formation, there is no source for metals, which are the most important gas cooling agents for star formation.  Once the photoionizing background is turned on, the halo gas is further heated and the gas becomes even more difficult to cool.  The result is an entirely dark halo at $z=0$.

\subsubsection{Simulations with alternate UV backgrounds}
\label{sec:uvsummary}

\begin{figure*}
\centering
\includegraphics[width=170mm]{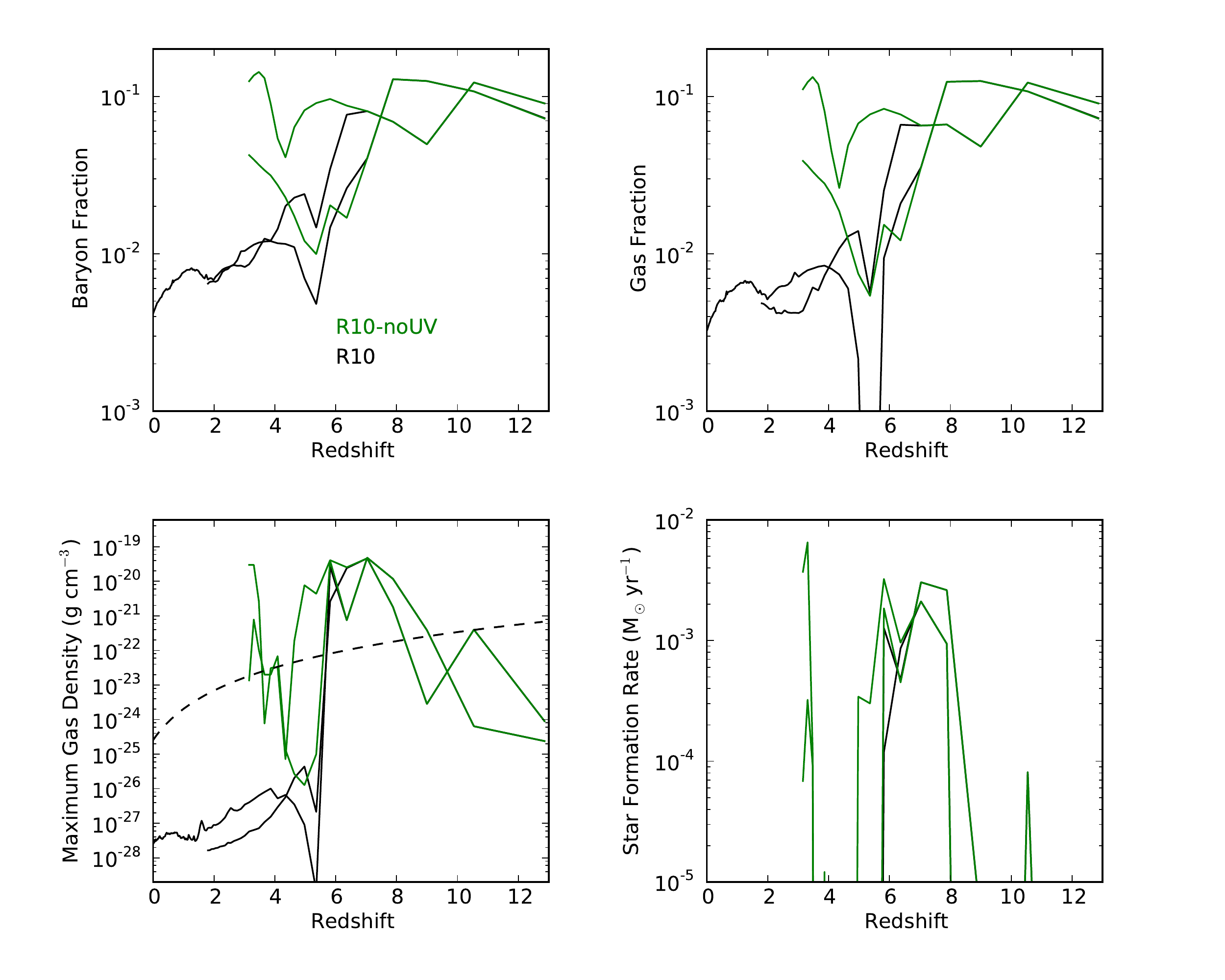}
\caption{Evolution of properties for two massive progenitors in simulations R10 (black) and R10-noUV (green).  R10-noUV was not run past a redshift of 3.1.  The properties shown here are the halo baryon fraction within the $r_{200}$ radius (top left); the halo gas fraction within the $r_{200}$ radius (top right); the density of the densest cell within each halo (bottom left); and the stellar mass within the $r_{200}$ radius for each halo (bottom right).  The density threshold for star formation is shown as a dashed line in the bottom left panel showing the peak gas density.}
\label{fig:evo_uvbg}
\end{figure*}

Our canonical simulations demonstrate that the UV background has an important effect on the evolution of the halo gas.  We observe that the gas fractions of progenitor haloes drop dramatically at the redshift of full-strength reionization in both our early reionization simulation (R10-earlyUV) and our late reionization simulation (R10).  We also observe gross differences in the properties of the final stellar component between these two simulations.  The total stellar mass in the final halo differs by an order of magnitude between the two runs and the star-formation history is more bursty and truncated in R10-earlyUV.  The median stellar metallicity is much lower in R10-earlyUV (0.06 \Zsun versus 0.51 \Zsun), a consequence of the lack of late and extended star-formation bursts.

To further explore the effect of the UV heating on the halo's evolution, we conducted a simulation without the photoionizing and photodissociating UV fields, R10-noUV.  This simulation was only run to a redshift of 3.1, but interesting comparisons can still be made.  Figure \ref{fig:evo_uvbg} shows the evolution of a variety of gas properties for the two massive progenitors of our final halo in R10-noUV (we show only two haloes for clarity).  There are some striking similarities and differences in the evolution of these haloes between R10 and R10-noUV.  We see dips in the gas fractions of haloes in R10-noUV, some quite dramatic, but the gas fraction is able to rebound in both massive progenitors.  By $z = 3.1$, when we stop R10-noUV, there is an order of magnitude difference in baryon and gas fractions for both haloes between the two simulations.  The baryon fraction in one halo has rebounded to its level at $z=12$.  

The dense gas exhibits similar behavior.  The peak gas density does drop considerably between redshifts of 6 and 4, but the gas is able to re-cool and condense to a level where it is able to form stars.  This produces an uptick in the star formation rate and several haloes are forming stars at $z=3.1$ when the simulation was ended.

\begin{figure*}
\centering
\includegraphics[width=170mm]{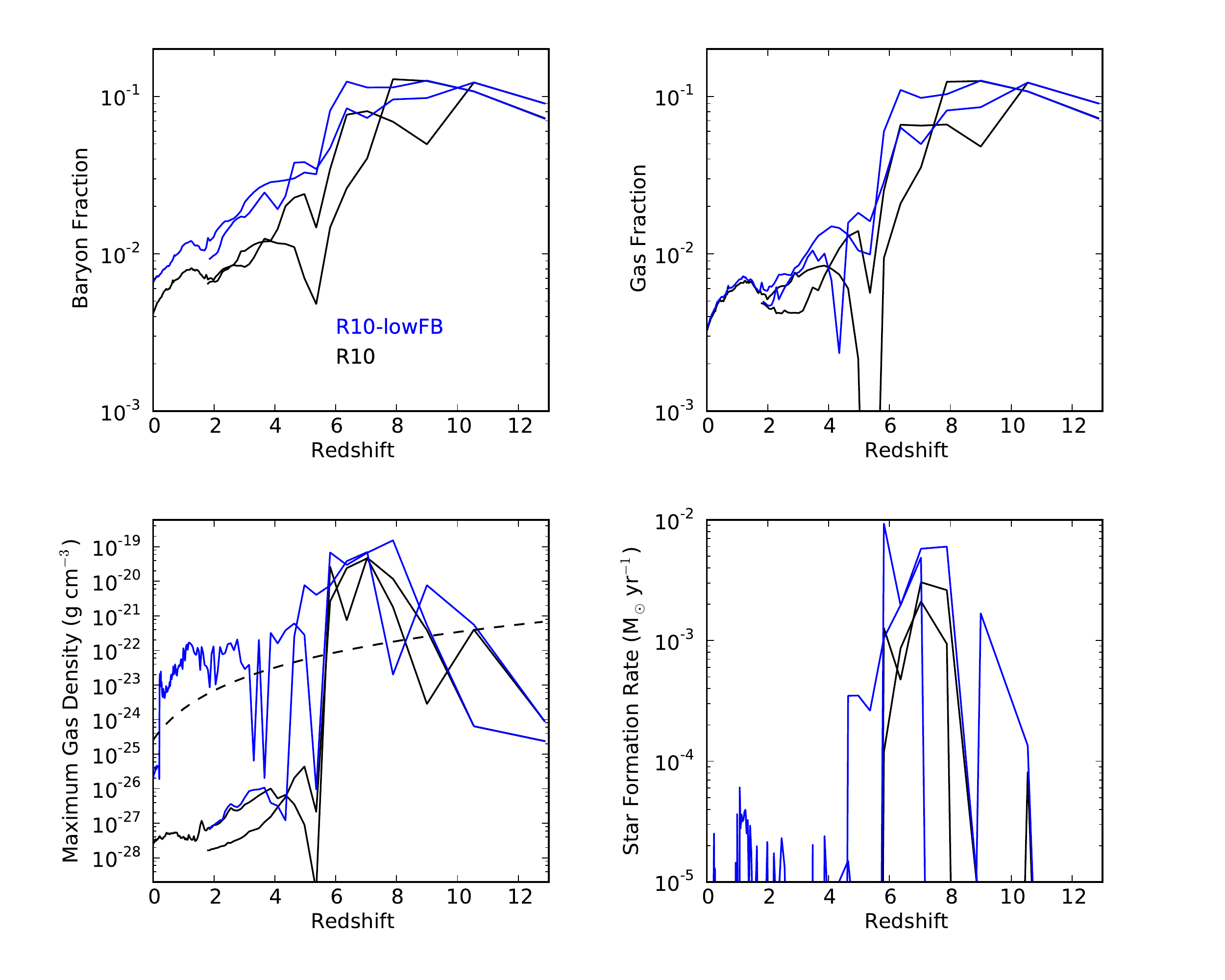}
\caption{Evolution of properties for two massive progenitors in simulations R10 (black) and R10-lowFB (blue).  The properties depicted are the same as those shown in Figure \ref{fig:evo_uvbg}.}
\label{fig:low_feedback}
\end{figure*}

The dips seen in the gas fraction and peak gas density in R10-noUV are most likely due to the effect of supernova feedback.  In the absence of UV heating, supernovae act as the main regulator of the star formation rate in this run.  The degree of supernova heating scales with the star formation rate and once the rate drops, gas is able to both re-accrete onto the halo and re-cool in its centre.  This produces an increase in the star formation rate.  We did not run this simulation beyond $z=3.1$, so it is impossible to say whether this cyclical process will continue or whether the constant bursts of star formation will produce an overall net decline in the gas fraction or peak gas density.

These simulations indicate that the UV background plays a crucial role in regulating the gas fractions of the progenitor haloes.  In our model, it produces a sharp drop in the gas fraction in all haloes.  This drop leaves an imprint on the star formation history inferred from the final stellar population of our target halo (the solid black lines in the bottom two panels of Figure \ref{fig:peak_density_cell_properties}).  This imprint is particularly evident because our UV backgrounds are spatially uniform, so all progenitors are affected with the same level of flux, and all the progenitor haloes are in the mass range sensitive to the UV backgrounds.  

\subsubsection{Simulations with alternate supernova feedback levels}
\label{sec:alternate_feedback}

\begin{figure*}
\centering
\includegraphics[width=170mm]{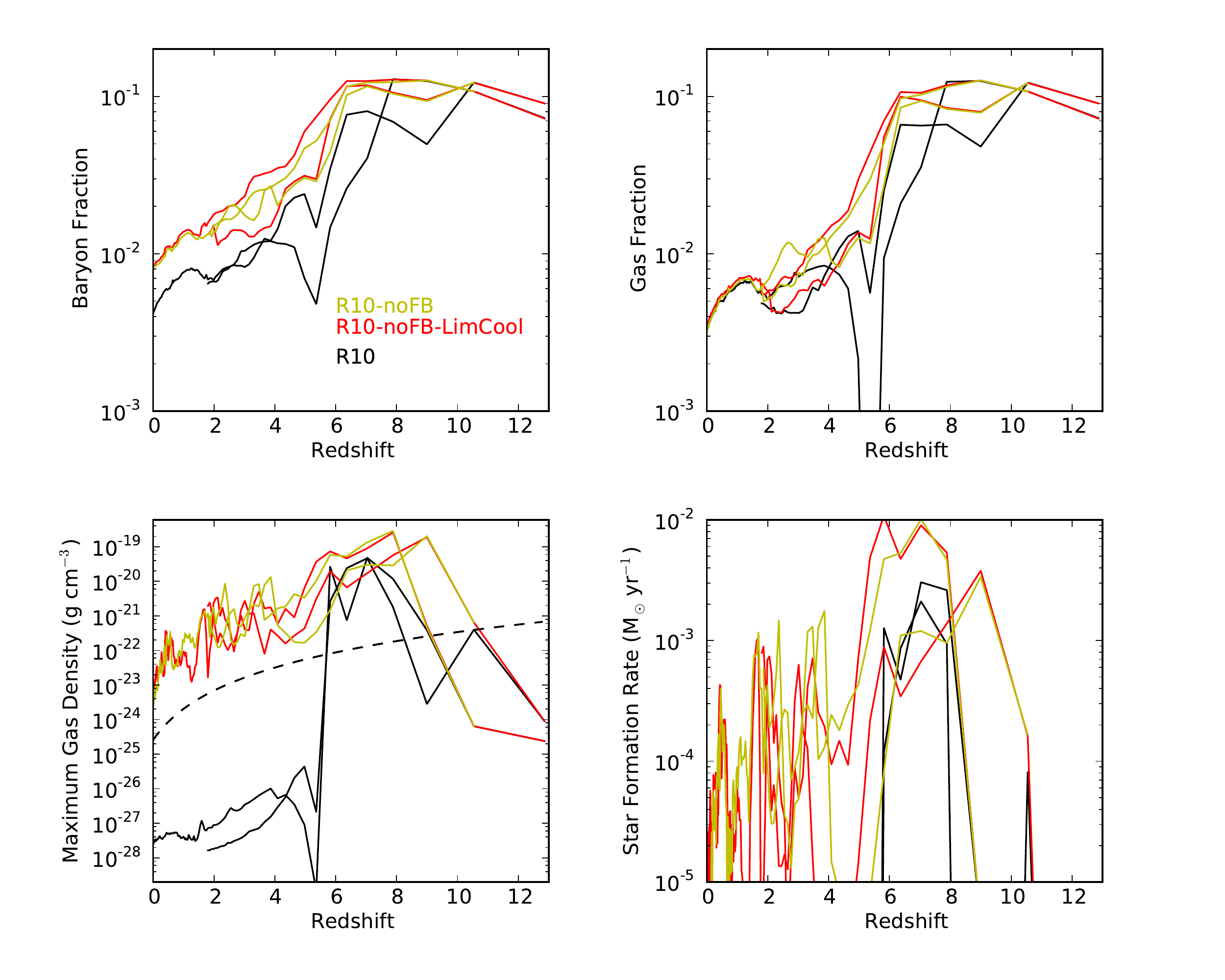}
\caption{Evolution of properties for two massive progenitors in simulations R10 (black), R10-noFB (yellow) and R10-noFB-LimCool(red).  The properties depicted are the same as those shown in Figure \ref{fig:evo_uvbg}.}
\label{fig:zero_feedback}
\end{figure*}

We also conducted three simulations exploring the effect of changing the strength of supernova feedback in our model.  In most of our simulations, including R10, star particles return $3.7 \times 10^{-6}$ of their rest mass energy to thermal gas energy.  In simulation R10-lowFB, we have set this fraction to be $10^{-6}$.  In simulations R10-noFB and R10-noFB-LimCool, star particles return zero thermal energy to the gas through supernovae, however, star particles do return metals to the gas in the same way as in R10 and R10-lowFB.

Properties for the two most massive progenitor haloes in simulation R10-lowFB are shown in Figure~\ref{fig:low_feedback}.  The gas fractions for these two haloes behave in nearly the same fashion as they do in simulation R10.  There appear to be smaller declines in the gas fraction prior to $z=6$, which must be due to the reduced impact of supernova heating.  Curiously, while the gas fraction declines in the same way and to the same levels as in R10, the peak gas density within each halo behaves differently.  There are sharp declines in the peak gas density between redshifts 6 and 4, but the peak gas density is then able to rise significantly in one of the progenitors, up to the level required for star formation.  

This difference in the evolution of dense gas results in steady, low-level star formation in R10-lowFB down to low redshift, unlike in R10.  The final stellar mass in R10-lowFB is $5.03 \times 10^6$ \Msun, almost five times the stellar mass found in R10.  The stellar component is also more metal rich than in R10: the log of the mean and median stellar metallicities is 0.30 and 0.10 respectively, more than 0.4 dex higher than in R10.  The longer star formation history, higher stellar mass and higher stellar metallicities are due to the decreased ability of supernova feedback to destroy self-shielded clumps and drive metals from star-forming gas in this simulation.

Star particles in our model impact the thermal state of the gas in two ways: first, by heating it through supernova feedback, and second, by injecting metals that act as coolants.  In simulations R10-noFB and R10-noFB-LimCool we have completely eliminated thermal feedback while maintaining the metal feedback.  Unsurprisingly, we see metals collecting in the centres of haloes to a much greater degree than in R10, making it unclear if changes in the no-feedback runs are due to a lack of energetic input, or an increase in the metal cooling rate.

To differentiate between these causes, in run R10-noFB-LimCool we capped the effective metallicity used for cooling at 0.1 \Zsun.  Figure~\ref{fig:zero_feedback} shows that the differences between R10-noFB and R10-noFB-LimCool are minimal; therefore the differences seen between these two simulations and R10 are due to a lack of heating by supernovae, not to enhanced cooling from the collected metals.  

The gas fractions of haloes in R10-noFB and R10-noFB-LimCool decline in almost the same way as in R10 after the introduction of the UV fields.  They are slightly elevated down to $z \sim 2$, as compared to the gas fractions in R10, but this may be due to their higher gas fractions at $z=6$ when the ionizing background reaches its full strength.  As in R10-lowFB, the overall decline in the gas fraction does not correlate with an overall decline in the peak gas density.  The peak gas density in both massive progenitors dips at $z=6$, but it does not fall below the threshold density for star formation, and we see that a low level of star formation persists in both progenitors down to $z=0$.

In all four simulations that explore different levels of supernova feedback (R10, R10-lowFB, R10-noFB and R10-noFB-LimCool), the final gas fraction is extremely low and virtually identical.  Comparing these results with those from R10-noUV, we can conclude that the factor that mainly determines the gas fraction of the halo is the UV background.  Supernova feedback does play an important role in regulating the dense gas component, from which stars are formed.  UV heating can produce an initial, sharp suppression of the peak gas density, however it is supernova heating that appears to determine whether it remains suppressed.

\section{Discussion}
\label{discussion}

We have explored a variety of physical effects in a series of high-resolution, hydrodynamical, cosmological simulations.  Our goals were to better understand the physics that regulate star formation in a dwarf halo comparable in mass to the dwarf satellite galaxies seen in the Local Group, as well as to test current feedback models in such systems.  We found that the combination of supernova feedback and reionization causes a large suppression of the star formation rates in these haloes.  Reionization appears to regulate the overall gas budget within the halo.  Supernova feedback seems to more directly affect the dense gas that is available for star formation.  Of course, these heating sources are not strictly exclusive in their effects; supernova feedback can have a quasi-periodic effect on the gas fraction and reionization can cause a dip in the peak gas density.  However, the quantities affected are typically able to recover in the absence of their primary regulator.

The reason for this dichotomy between reionization and supernova feedback is primarily their different physical locations within the halo.  The UV backgrounds we implement are uniform and are attenuated by our self-shielding prescriptions in dense gas at halo centres.  At the onset of reionization, the ionizing background is able to produce a pressure imbalance nearly everywhere in the halo except the very centre.  A large fraction of gas is therefore driven out of the halo in photoevaporative winds, while the central dense gas can remain intact for some time \citep{alvarez06, gnedin06, whalen08} 

Thermal energy from supernova feedback is only injected into the single cell containing a young star particle in our model.  Since the star particles tend to form and remain centrally located in haloes throughout the simulation, the energy they inject tends to affect central dense gas.   We do see evidence that supernovae in our model affect regions outside the central dense gas in the form of supernovae driven outflows which enrich the IGM with metals, as seen in Figure~\ref{fig:image_panel_plot} and discussed more below.  It is unclear, however, to what extent our supernova feedback model accurately captures the heating of the ISM.  While the simulations that we have run with little or no thermal feedback seem to indicate that our feedback prescription is key in expelling dense gas and truncating the star formation history, we will explore this issue in more detail later in this section. 

In the following sections, we will first compare our results to observed properties of low mass dwarf systems, then we will examine the effect of supernova heating in more detail, followed by discussions of spatial resolution  and metal dispersal, and ending with a discussion dark matter properties and neglected physics.

\subsection{Observable properties}

In this section we compare our models to observations of several types of low-luminosity dwarf galaxies.  This comparison is important because understanding the ways in which our simulations fail to produce realistic galaxies can provide important insights into the specific ways in which our model prescriptions fail.  We can consider two classes of dwarf galaxies for comparison: mass analogues and environmental analogues.  Our simulated halos have stellar masses and velocity dispersions similar to those of dSphs found near the Milky Way.  Unfortunately, their proximity to a large galaxy does not make them suitable environmental analogues; our simulated halo is 2.9 Mpc from a Milky Way sized halo.  Dwarf galaxies have been observed in underdense environments, however, most of these are dIrrs that are more luminous than both the typical Milky Way dSphs and our simulated haloes \citep{weisz11a}.  A handful of dSphs are known in low-density environments \citep[e.g.][]{makarov12}, however while some of these systems may be relics of reionization \citep{monelli10}, others likely underwent pericentric passages with a large host galaxy \citep{teyssier12}.  Many of the lowest luminosity dwarfs discovered around the Milky Way have been found in the SDSS, for which the distance limit for detection has been demonstrated to be less than the Milky Way virial radius \citep{koposov08}.  It is therefore possible that similar low-luminosity dwarfs await detection at greater distances in low-density environments.  

\begin{figure}
\centering
\includegraphics[width=90mm]{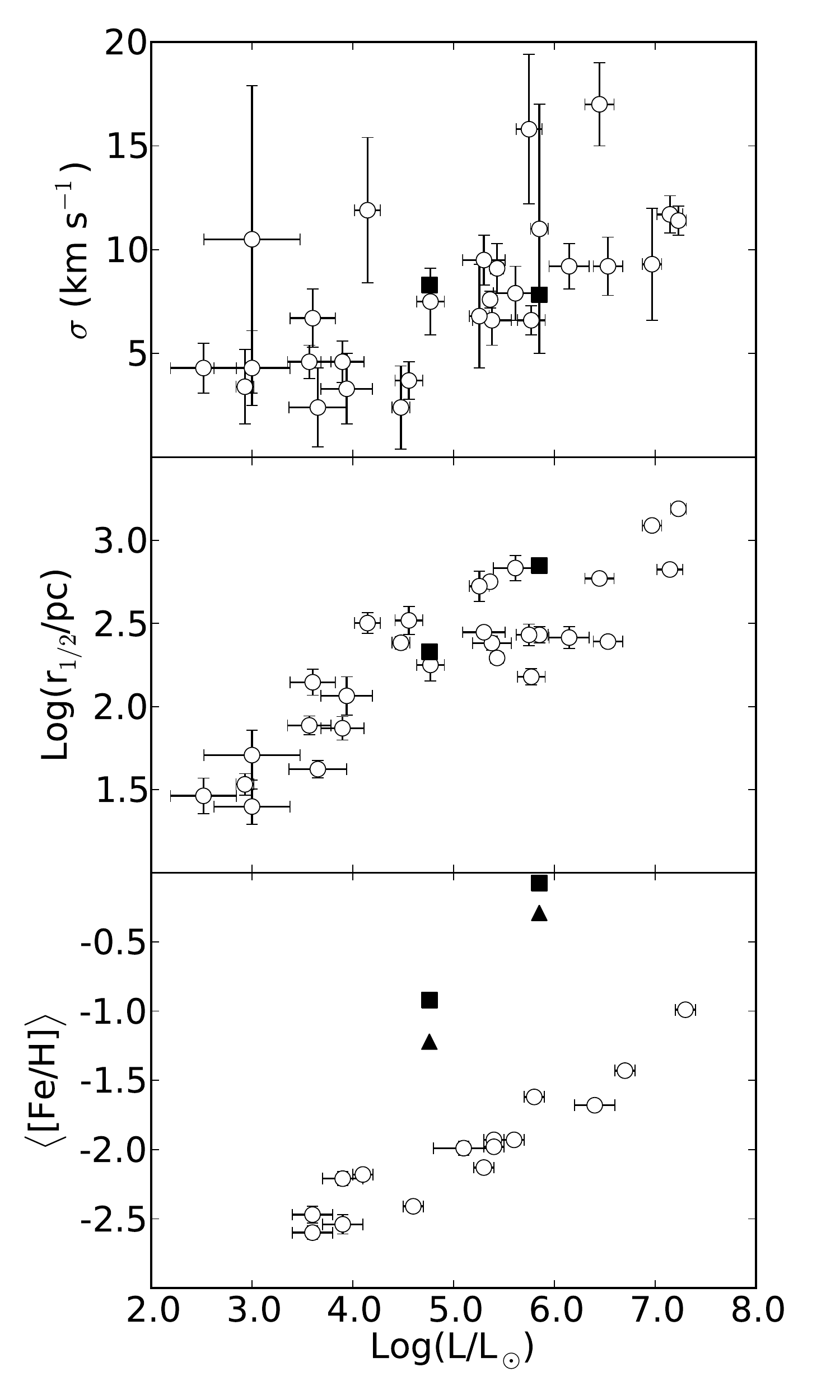}
\caption[Caption]{Comparison between the observational properties of Local Group dSphs and the simulated properties of the final haloes produced in R10 and R10-earlyUV.  All quantities are plotted against total V band luminosity.  V band luminosities are estimated for our simulated haloes as described in section \ref{sec:stellar_mass}.  \textit{Top panel:} Velocity dispersions for 28 Local Group dSphs collected by \citet{walker09b}\footnotemark, but including the velocity dispersion for the dynamically cold component of Bootes I measured by \citet{koposov11} (open circles); and the three-dimensional velocity dispersions of star particles within the half-stellar mass radii of R10 and R10-earlyUV (solid squares).  \textit{Middle panel:} The half-light radii for the same sample of dSphs (open circles) and the half stellar mass radii for the final haloes in R10 and R10-earlyUV (solid squares).  \textit{Bottom panel:} The mean stellar metallicities for 15 Milky Way dSphs measured by \citet{kirby08} and \citet{kirby11a} (open circles), the mean (solid squares) and median (solid triangles) star particle metallicities for the final haloes in R10 and R10-earlyUV. }
\label{fig:obs_comp}
\end{figure}

As shown in Table~\ref{tab:final_properties}, our canonical simulations R10 and R10-earlyUV produce final halos with dynamical masses within 300 pc close to $10^7$ \Msun, similar to those derived for Milky Way dSphs \citep{strigari08}.  The final velocity dispersion of star particles within the radius enclosing half the stellar mass, $r_{1/2}$, is roughly 8 km s$^{-1}$ in both simulations, making them dynamical analogues of Milky Way dSphs as shown in Figure \ref{fig:obs_comp}.  In the following subsections we discuss other observational comparisons.

\subsubsection{Total stellar mass}
\label{sec:stellar_mass}

The final stellar masses produced in our simulated halo are $1.43\times10^6$ \Msun\ in R10 and $1.16\times10^5$ \Msun\ in R10-earlyUV.  Assuming a stellar mass-to-light ratio in the V band of 2 \Msun/\Lsun\ for our (old) stellar population \citep{kruijssen09}, the haloes in R10 and R10-earlyUV would have V band luminosities of approximately $7.15 \times 10^5$ \Lsun\ and $5.8 \times 10^4$ \Lsun \ respectively.  These estimates place the halo in R10-earlyUV at the upper luminosity end of the ultrafaint dSphs and R10 among the lower luminosity classical dSphs \citep{walker09b,simongeha07}, as shown in Figure \ref{fig:obs_comp}.  If we compare our estimated luminosities to the haloes' total gravitational masses, these haloes have mass-to-light ratios within $r_{1/2}$ (the radius enclosing half the stellar mass) of 85 \Msun/\Lsun\ for R10 and 133 \Msun/\Lsun\ for R10-earlyUV.  The spatial extent of the final stellar component is also comparable to real systems; Figure \ref{fig:obs_comp} shows that the half stellar mass radii for the final haloes in R10 and R10-earlyUV are similar to the half-light radii of dSphs with similar luminosities.  

In other highly resolved cosmological calculations of low-mass dwarf haloes, the total stellar mass is generally similar to what we have found.  \citet{sawala10} found, for their highest resolution model of a similar halo mass, a simulated stellar mass of $7.35\times10^6$ \Msun.  Recent work by \citet{governato12} also probed halo masses close to $10^9$ \Msun, and while the stellar content is not the focus of their study, their models appear to produce only $\sim5\times10^4$ \Msun\ of star particles for haloes close to this mass.  One of the many differences between these two studies is the time of reionization; \citet{sawala10} reionize the universe at $z\sim6$ and \citet{governato12} do so much earlier at $z\sim9$.  Considering this, the results of \citet{sawala10} are most analogous to our simulation R10, which has a final stellar mass of $1.43\times10^6$ \Msun, and the results of \citet{governato12} are most analogous to simulation R10-earlyUV which has a final stellar mass of $1.16\times10^5$ \Msun.  In this light, these results appear to be roughly consistent with each other (and with the idea that later reionization leads to larger stellar masses), although differences in final stellar mass between analogous haloes can approach an order of magnitude.

\footnotetext{These data were taken by \citet{irwin95}; \citet{walker09c}; \citet{martin08}; \citet{walker07}; \citet{mateo08}; \citet{koch07}; \citet{martin07}; \citet{koch09}; \citet{simongeha07}; \citet{aden09}; \citet{belokurov08}; \citet{walker09a}; \citet{irwin07}; \citet{geha09}; \citet{belokurov09}; \citet{mcconnachie06}; \citet{cote99}; \citet{chapman05}; \citet{ibata07}; \citet{letarte09}; \citet{lewis07}; \citet{ibata07}; \citet{majewski03}; \citet{saviane96}; \citet{fraternali09}}
\subsubsection{Metallicity}

Milky Way dwarfs are observed to lie on a fairly tight luminosity-metallicity relation \citep{kirby11a}, shown in Figure \ref{fig:obs_comp}, which may extend beyond the Milky Way \citep{kirby12}.  Figure \ref{fig:obs_comp} shows the metallicity-luminosity relation found for Milky Way dSphs.  Star particles in R10 have a mean metallicity of 0.84 \Zsun\ and a median metallicity of 0.51 \Zsun; in R10-earlyUV they have a mean metallicity of 0.12 \Zsun\ and a median metallicity of 0.06 \Zsun.  If we make the estimate that [Fe/H]~$\sim$~log(Z/\Zsun) \citep[e.g.][]{woo08}, then these values translate to mean and median [Fe/H] values of -0.076 and -0.29 dex for R10 and mean and median [Fe/H] values of -0.92 and -1.2 dex for R10-earlyUV.  Our assumption that [Fe/H]~$\sim$~log(Z/\Zsun) is a somewhat flawed one given the short duration of star formation in our progenitor haloes, and the fact that we do not properly model contributions from SNIa.  We would expect this short duration to result in elevated values of [$\alpha$/Fe] for their stellar populations \citep{tinsley79}.  This has the potential to decrease our estimated [Fe/H] by a few fractions of a dex, but we are interested in understanding differences in [Fe/H] of a dex or more.

Based on our estimated luminosities calculated in Section \ref{sec:stellar_mass} and according to the dSph luminosity-metallicity relation of \citet{kirby11a}, star particles in the final halo in R10 should have an average [Fe/H] of -1.76 dex ($1.7\times10^{-2}$ \Zsun) and -2.09 dex ($8.1\times10^{-3}$ \Zsun) in R10-earlyUV.  Figure \ref{fig:obs_comp} shows this discrepancy, which is over 1.5 dex for R10 and over 1 dex for R10-earlyUV.  This 1-2 dex difference between observed systems and our simulations has important implications for the utility of the feedback model we use.  It appears that while our adopted feedback model effectively suppresses the formation of dense gas and stars, it does not drive sufficient metal ejection from the ISM.

\subsubsection{Gas content and star formation history}

\begin{figure}
\centering
\includegraphics[width=90mm]{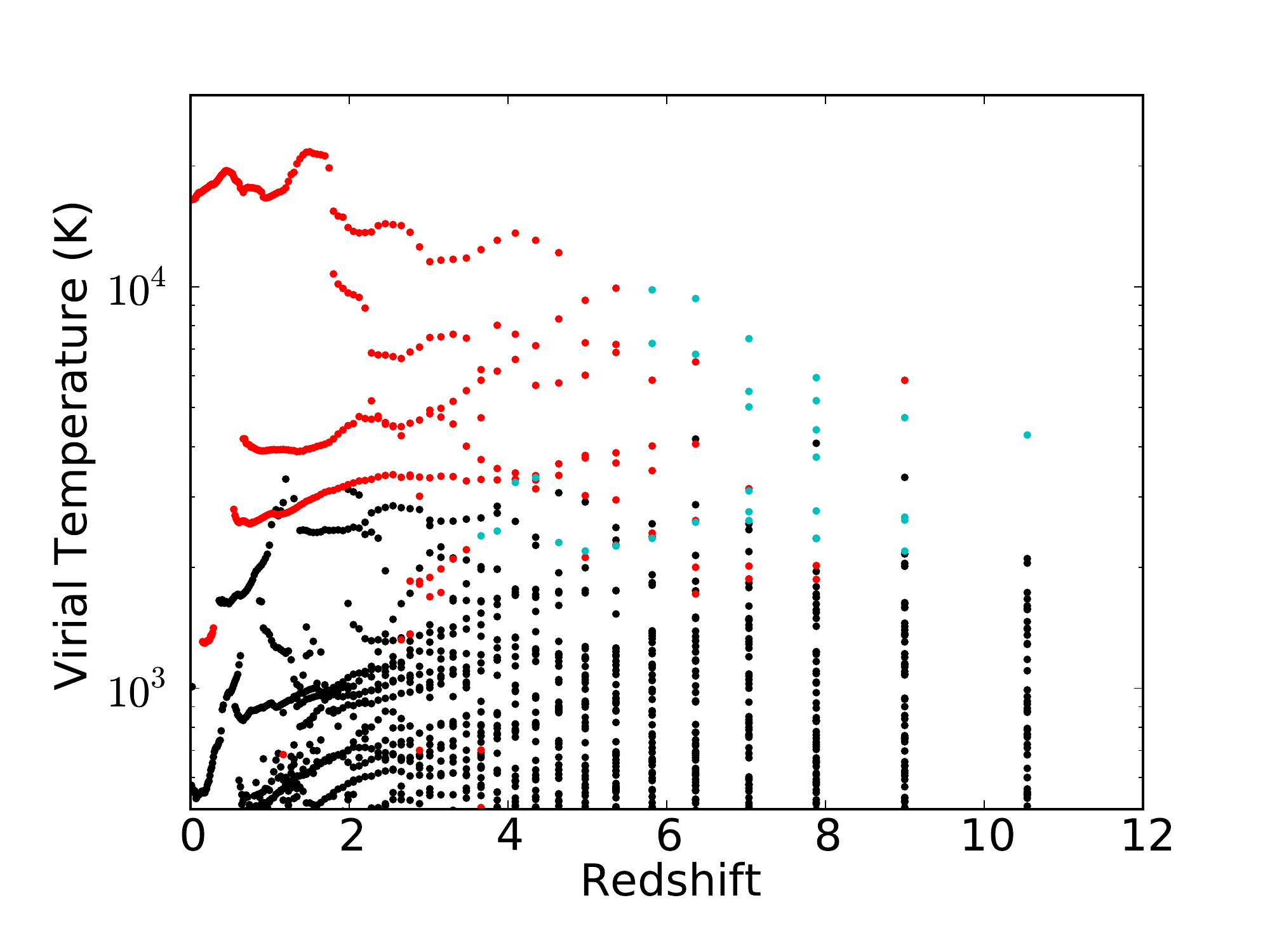}
\caption{Virial temperature versus redshift for progenitor haloes that are constituted of at least 50\% final target halo dark matter particles.  Haloes plotted in black contain no star particles; haloes plotted in cyan have formed star particles within the last 20 Myrs; and haloes plotted in red contain star particles, but have not formed star particles within the last 20 Myrs.}
\label{fig:tvir}
\end{figure}

Our simulated halo in R10 and R10-earlyUV, has a final gas fraction that is very suppressed; the total mass in gas is less than 1\% of the total halo mass, well below the cosmic baryon fraction (see Table \ref{tab:final_properties}).  As we have discussed in Section \ref{sec:detailed_evo}, this remaining halo gas is very diffuse and highly ionized with a temperature of a few factors of $10^4$ K (see the $z=0$ panel of Figure \ref{fig:phaseplots}) and would not be directly observable in a real system.  

Many low-luminosity dSphs in the Milky Way are very gas poor with little or no H I, but (unlike our simulated system) some of the more distant Milky Way dwarfs can be quite H I rich, for example Leo T \citep{ryanweber08}.  Our simulated halo is isolated and has had no interactions with the hot halo of a massive galaxy during its formation.  We might, therefore, expect it to be more analogous in its gas content to the more distant Local Group dwarfs.  The observed strong correlation between environment and gas content for Local Group dwarfs \citep{grcevich09} suggests that their gas content may be due to environmental effects, which this study does not explore since we only simulate one environment.

Reionization appears to be the mechanism that removes most of the gas mass in our models and the only simulation with evidence for late time gas accretion is R10-noUV, which had no UV background.  From this simulation, we conclude that the lack of reaccretion of gas onto our simulated halo is due to its low virial temperature relative to the warm temperature of the IGM after reionization, a consequence of the halo's low mass ($T_{vir} \propto M^{2/3}$).  It does not appear that supernovae are a factor in preventing the reaccretion of gas, as the suppression of gas fractions is remarkably similar in our simulations with different levels of supernova feedback.

Figure \ref{fig:tvir} shows that at late times the main progenitor haloes and massive subhaloes in R10 all have virial temperatures well below the temperature of the IGM (a few factors of $10^4$ K).  This plot also shows that, prior to reionization, with a few exceptions, progenitor haloes generally need to have virial temperatures above 2000 K in order to form stars, due to the inefficiency of H$_2$ cooling, see \citet{tegmark97}.  After reionization, most progenitor haloes stop forming stars.  As we discussed in Section \ref{MainResults}, this suppression of star formation was due to a combination of the large scale photoevaporation of gas from progenitor haloes and the effect of supernova heating on the remaining self-shielded clumps of dense gas.  The net effect is consistent with the idea that reionization introduces a filtering mass or filtering virial temperature, below which star formation is truncated \citep[e.g.][]{gnedin00}. 

Observations of dwarfs in the present epoch cannot tell us the complete evolution of gas in these systems, but their star formation histories derived from color-magnitude diagrams do give us the history of dense star-forming gas.  We find that our simulated star formation histories in R10 and R10-earlyUV are sharply truncated after reionization.  R10 has the most extended star formation history; however, all star formation in this model ends by a redshift of 3.5.  Milky Way dSphs that are closest in luminosity to R10 and R10-earlyUV have a range of velocity dispersions and star formation histories.  In general, the most truncated star formation histories tend to be found in systems with the lowest velocity dispersions ($\sim$ 3 km s$^{-1}$) \citep{brown12}, although recent analysis suggests at least one system with a luminosity and velocity dispersion comparable to our simulated haloes, Canes Venatici I, has only ancient stellar populations \citep{okamoto12}.  In general, however, most of the Milky Way dwarfs contain some intermediate-age stellar population \citep{tolstoy09} which we do not see in our simulations.  

\subsubsection{Implications of observational comparisons}

Given the difficulties in determining the correct observational comparison sample as discussed in the first paragraph of this section, what can we conclude about the discrepancies between our simulated haloes and observed Local Group dwarf systems, especially in regards to the lack of an intermediate age (i.e. 1-8 Gyr old, see \citet{tolstoy09}) stellar population?  
The star-forming state of our simulation appears to be heavily dependent on the metagalactic UV background we implement and this is relatively well understood for $z < 6$.  The lack of dense, star-forming gas in our simulations at late times implies one of four things: one, either the implementations of the metagalactic UV background or the local shielding are inadequate and a larger reservoir of gas should survive reionization in these systems; two, our supernova feedback prescription is too effective in destroying dense gas that survives reionization; three, the handful of gas rich, but low luminosity, dwarfs (such as Leo T) observed at large galactic radii have virial temperatures above that of the IGM (and above the halo we simulate) which allows them to accrete gas at late times; or four, our halo merger tree is atypical for this halo mass or is atypical for dwarfs that end up in dense environments.  

The first scenario is possible and more sophisticated methods may be used to simulate ionization and shielding, as we will discuss in Section \ref{sec:neg_phys}.  However, self-shielding would have to be many times more effective than in our current model to significantly impact the gas fractions.   The second scenario also seems unlikely, since the discrepancy in stellar metallicity between our simulations and observed systems indicates that our feedback prescription is not effective \textit{enough}.  Indeed, in R10-lowFB, we found with weaker thermal feedback that star formation did continue down to low redshift, but star particles were even more metal rich.  If the third scenario is true, then we would predict undiscovered dwarfs with the luminosities, star formation histories and gas content similar to Milky Way dSphs in the field.  The fourth possibility is that our halo merger tree is different from dwarfs found in denser environments.  Although we do not study this directly, this possibility is plausible.  It is certainly the case that the typical mass of progenitor haloes at the time of reionization greatly impacts the final properties of the system and this typical mass may be different for different merger histories.  It is interesting to note, however, that analysis of more isolated, but more luminous, dwarfs in the local volume indicates general consistency in star formation histories between these systems and Milky Way dwarfs \citep{weisz11b}

\subsection{Supernova feedback model}
\label{sec:feedback_model}

In this sub-section, we first compare suppression of star formation due to supernova feedback in our simulations to previous work.  Then we explore the effectiveness of our model with simple analytic arguments, and finally suggest improvements.

Our supernova model involves a simple injection of thermal energy into the ISM over a few dynamical times.  For our canonical models we chose the level of thermal energy to inject into the gas to be $10^{51}$ ergs per 150 \Msun\ of star particles.   As discussed in Section~\ref{sec:alternate_feedback}, we find that lowering the feedback energy, as in simulations R10-lowFB, produces a more extended star formation history with star formation continuing past $z=1$.  The suppressed star formation histories in R10 and R10-noUV are generally consistent with other cosmological simulations probing halo masses of $10^9$ \Msun.  \citet{governato12} found star formation to be strongly suppressed in haloes with viral temperatures below that of the IGM and did not see the formation of central dark matter cores in haloes of this mass, which were produced in higher mass haloes by repeated starbursts extending down to late times.  \citet{sawala10} found star formation lasted approximately 2.62 Gyrs in their highest resolution simulation closest in mass to our simulated halo, a somewhat longer star formation epoch than found in R10.

\begin{figure}
\includegraphics[width=94mm]{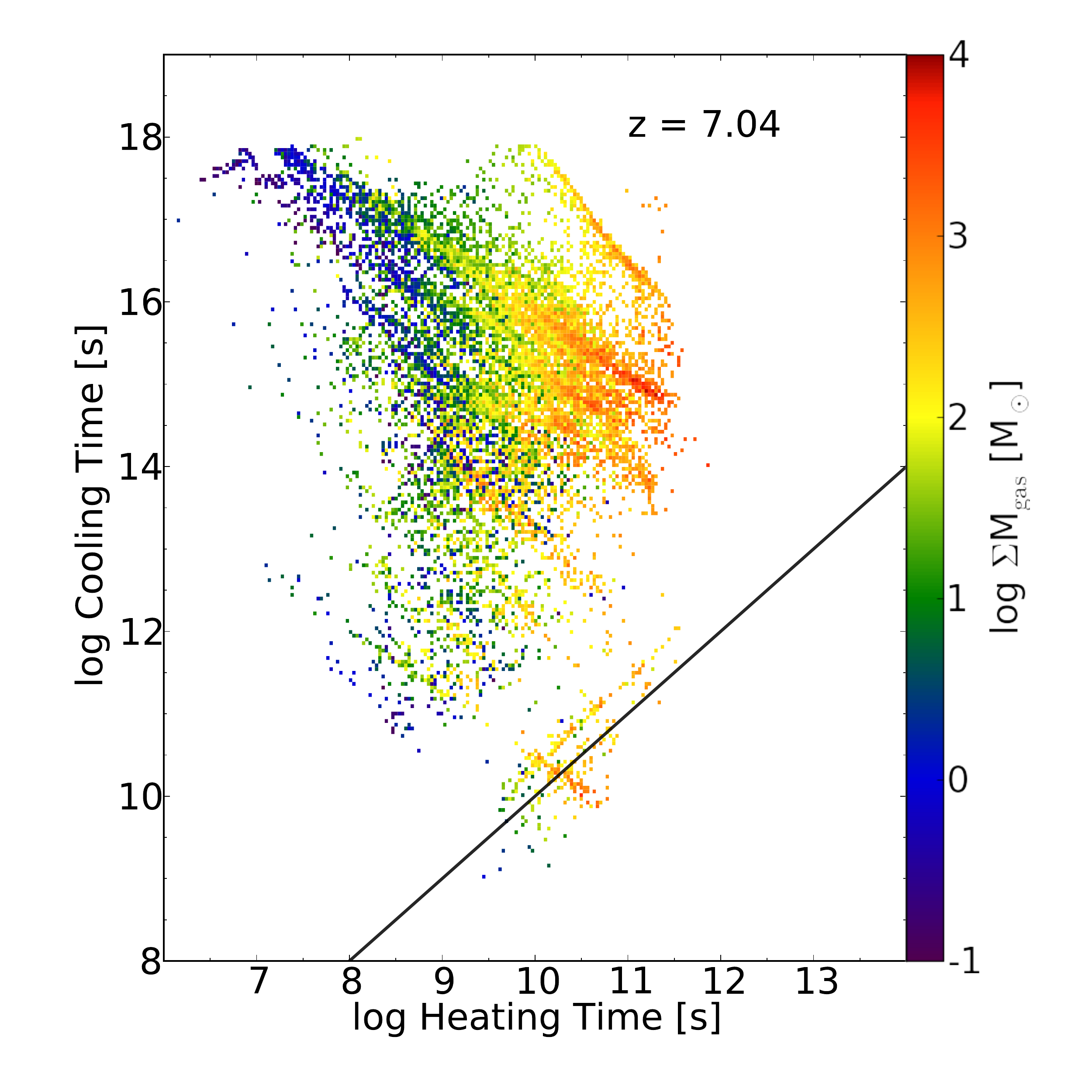}
\caption{Cooling time versus estimated heating time at $z=7.04$ for cells within $r_{200}$ for the massive progenitor whose phase evolution is shown in Figure~\ref{fig:phaseplots}.  The line where the cooling time equals the heating time is shown in black.  The colour of cells shows the cumulative sum of mass within the corresponding cooling time-heating time bin.}
\label{fig:heat_cool_phase}
\end{figure}

While a lower level of energy conversion may be indicated by the lack of late time star formation in our simulated halo, the metallicity of the star particles produced in R10 and R10-earlyUV exceeds that observed, arguing in the opposite direction.  To better understand how supernova heating occurs in our model, we examine heating and cooling timescales for each cell.  We estimate the heating time of a cell to be the ratio of the cell's thermal energy to the average supernova heating rate, assuming that exactly one star particle is heating that cell, spread out over the typical timescale we adopt in the code:

\begin{equation}
\label{eq:theat}
t_{heat} = \frac{3}{2} \frac{m_{cell}kT_{cell}}{\mu_{cell} m_{H}} \frac{t_f}{e_{SN} M_* c^2} ,
\end{equation}

\noindent
where $e_{SN} = 3.7\times 10^{-6}$ is the fraction of its rest mass energy a star particle returns to the gas; $t_f$ is the time over which a star particle returns its supernova energy to the ISM in our model, which we take to be 10 Myr; and $M_* = 100$ \Msun, a typical star particle mass in our simulations.  We computed $t_{heat}$ for every cell within r$_{200}$ and compare it to the cooling time of that cell in Figure~\ref{fig:heat_cool_phase}, computed at $z=7.04$.  To be clear, the cells for which we have estimated this heating time are not necessarily being heated by a star particle at the time shown; we are simply estimating a timescale for the response to such heating.

From Figure \ref{fig:heat_cool_phase}, we see that for nearly all of the cells within the halo, the cooling time far exceeds the potential heating time.  However, we see that gas with the shortest cooling times lies along the line where heating time equals cooling time.  This gas is most likely the star forming gas since star particles tend to form in cells with lower cooling times (more precisely in cells where the cooling time is shorter than the dynamical time).  Star particles are more likely to heat this gas since young star particles do not wander too far from their formation sites.

\begin{figure*}
\centering
\includegraphics[width=170mm]{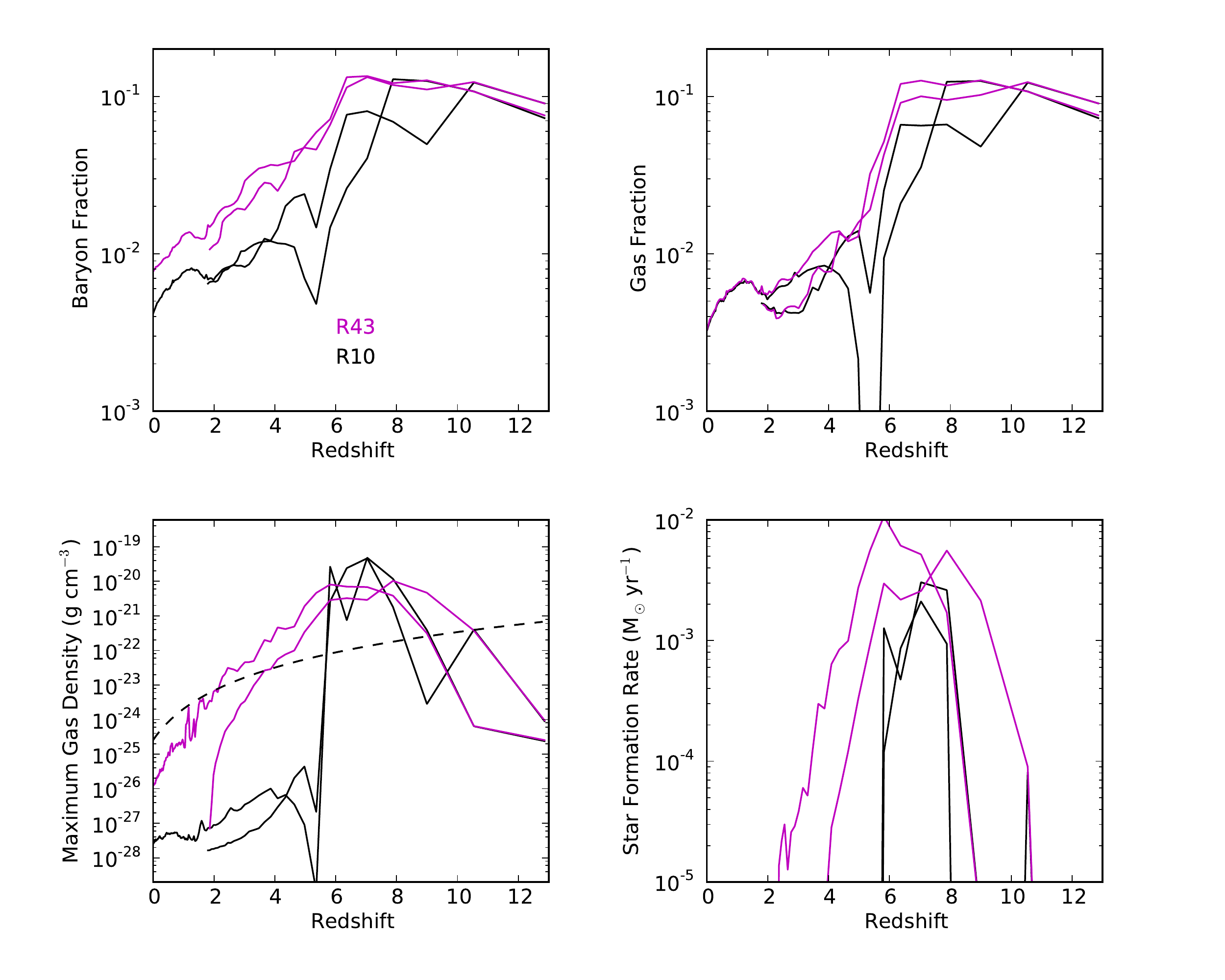}
\caption{Evolution of properties for two massive progenitors in the canonical simulations R10 (black) and the lower minimum resolution simulation R43 (magenta).  The properties depicted are the same as those shown in Figure \ref{fig:evo_uvbg}.}
\label{fig:resolution}
\end{figure*}

We interpret this result to mean that some energy produced by our feedback model is cooled in an unrealistic fashion.  The key issue is most likely the length of time over which supernova energy is injected into the ISM, i.e. the length of $t_f$.  Our feedback algorithm was motivated by the requirements of simulating larger mass halos and was designed to model the release of supernova energy from an entire stellar population that forms over an extended period of time (more than $10^7$ years).  A typical star particle in our simulations has a mass of 100 \Msun, which is approaching the mass of an individual massive star.  Extending the release of supernova energy over $10^7$ years is therefore not suitable for this smaller mass star particle which is producing about one supernova's worth of energy.  Reducing $t_f$ or even eliminating it and instantaneously injecting energy into the cell would be more appropriate.  Reducing $t_f$ would result in a shorter $t_{\rm heat}$ (eq.~\ref{eq:theat}), and so cooling would not be able to act before the injected feedback energy begins to operate.  Similar models have been explored in high-resolution local simulations of the ISM \citep{joung06}.  We found, however, that even with our high spatial resolution, injecting a supernova's worth of energy into one resolution element instantaneously proved to be computationally challenging and we are working to develop an improved algorithm.

In our current model implementation, energy that is able to escape the central regions with shorter cooling times will not be quickly lost since gas everywhere outside the central region has comparatively long cooling times.  Indeed in Figure \ref{fig:image_panel_plot} we do see evidence for supernova heated bubbles and metal pollution of the ISM and IGM.  Previously in Section~\ref{MainResults}, we had concluded that reionization determines the overall gas fraction of haloes and that dense gas is more directly affected by feedback.  This interpretation still largely holds, however, the effect on dense gas may be more pronounced than what we have observed in our simulations.  This would have the effect of enhancing the dominance of feedback over the UV background as the regulator of dense gas.

\subsection{Resolution effects}
\label{sec:resolution}

To explore the impact of spatial resolution, we examine simulation R43, which is a run with the same dark matter particle resolution and physics as R10, but with four times coarser spatial resolution.  R43 has a minimum cell length of 43 comoving parsecs.  Figure~\ref{fig:resolution} shows the evolution of relevant quantities for the two main star forming progenitors in R43 and Table \ref{tab:final_properties_r43} lists many of the final halo's properties.  The gas fractions of haloes in R43 evolve in a similar way to those in R10, with a large decline around $z = 6$.  The declines in gas fraction prior to reionization, however, are almost nonexistent.  These declines in R10 were due to supernova feedback and we see that coarser spatial resolution has decreased its effect.

The dense gas, which we found was more directly regulated by supernova feedback, also has very different behavior.  The dense gas in the two main progenitors does not rise to quite as high a peak density, but is able to survive longer because of the diminished impact of supernovae in the low resolution run.  Cells in R43 encompass 64 times the volume of cells in R10; therefore when a star particle injects energy into a cell in R43, the specific heating rate is approximately $1/64^{\rm{th}}$ of what it is in R10.  Dense gas is therefore able to survive longer.  The maximum density cell within each halo peaks at a somewhat higher density in R10 because we are able to better resolve the collapse of cold clouds.

The star formation events in R43 are of longer duration due to the enhanced survivability of dense gas.  Star formation continues in the main massive progenitor down to $z=2$.  This produces a final halo with a greater stellar mass ($6.82 \times 10^6$ \Msun).  This is somewhat counter to the expectation that increased resolution will produce more star formation (since we resolve denser gas), however this result makes sense in light of the resolution-dependent effectiveness of our supernova feedback model.

\begin{table}
\centering

  \caption{Summary of Final Halo Properties for Coarse Resolution Simulation R43. \label{tab:final_properties_r43}}
  \begin{tabular}{@{}cc}
  \hline
   & R43 \\
  \hline
 $M_{tot}/M_\odot$ & $1.54 \times 10^9$\\
$M_*/M_\odot$  & $6.82 \times 10^6$\\
$M_{gas}/M_{tot}$ & $3.28 \times 10^{-3}$\\
$r_{200}$ (kpc) & 23.7\\
$r_{1/2}$ (pc) & 326\\
$M_{1/2}/M_\odot$ & $1.56 \times 10^7$\\
$M_{300}/M_\odot$  & $1.42 \times 10^7$\\

$\sigma_{1/2}$ (km/s) & 8.56 \\

Log($\langle Z/Z_\odot \rangle$) (median)&  -0.1\\
Log($\langle Z/Z_\odot \rangle$) (mean) & 0.0\\
$\sigma_Z/Z_\odot$ & 0.83\\

 \hline
\end{tabular}

Note: The quantities presented in each row are the same as in Table \ref{tab:final_properties}. 

\end{table}

Two other high-resolution, cosmological simulation studies track the evolution of similar mass haloes to $z=0$: \citet{sawala10} and \citet{governato12}.  Both studies use a SPH code, so it is not possible to directly compare our resolutions in a consistent way.  However, both simulations are closer in spatial resolution to our simulation R43; \citet{sawala10} fix their softening length in such a way that their resolution is typically less than 100 pc and \citet{governato12} have a softening length of 64 pc for their highest resolution run.  In simulations of somewhat more massive haloes ($10^{10}$ \Msun), \citet{governato10} found that increasing resolution was key in reducing the central angular momentum of galactic disks that form in their haloes and attribute this difference to better resolving spatially separated clouds in their higher resolution simulations, which enhanced the effect of supernova feedback as a regulator of star formation.  They also found convergence in behavior between their medium resolution simulation (with a force resolution of 116 pc) and their high resolution simulation (with a force resolution of 86 pc), in contrast to the behavior of our simulations R10 and R43.  Therefore, while it appears that resolution in key in simulating star formation and supernova feedback in galaxy simulations, the exact mechanisms at play and the scales on which behavior converge are dependent on the physical prescriptions and hydrodynamical methods used.

Simulations R10 and R43 do not converge on a star-formation history or baryon fraction for our final halo, however we argue that many of the conclusions we have drawn from our higher resolution simulations hold.  Clearly, reionization and feedback operate in the same way in both simulations, where the overall gas fraction is regulated by reionization and the dense gas evolution is regulated by supernova feedback.  We do see an increasing effect of supernova feedback on the gas fraction with increasing resolution, so it is possible that increasing the resolution even more will have an even greater effect.  The effect of reionization, however, is unchanged and still quite dominant.  We may not have converged on a final stellar mass for this halo.  We therefore conclude that our measures of the final stellar mass are likely to be an upper limit. 

\subsection{Metal ejection}

\label{sec:metal_ejection}

We find that the main mechanism for gas ejection from these low mass haloes is photo-evaporative winds due to reionization.  It remains unclear, however, if this is also the mechanism for metal ejection.  We see evidence for metal rich supernova heated winds in Figure~\ref{fig:image_panel_plot}, but it also appears from Figure~\ref{fig:peak_density_cell_properties} that the metallicity of dark haloes increases after reionization.

We can answer this question by comparing simulations with and without supernova feedback and reionization: R10, one of our canonical runs that contains our complete physical model; R10-noFB-LimCool, which has no thermal supernova feedback; and R10-noUV, which has no UV background.  Figures~\ref{fig:evo_uvbg} and \ref{fig:zero_feedback} show that these three simulations have produced varying amounts of star particles over somewhat different star formation histories, but we can nevertheless draw robust conclusions.  

\begin{figure}
\centering
\includegraphics[width=60mm]{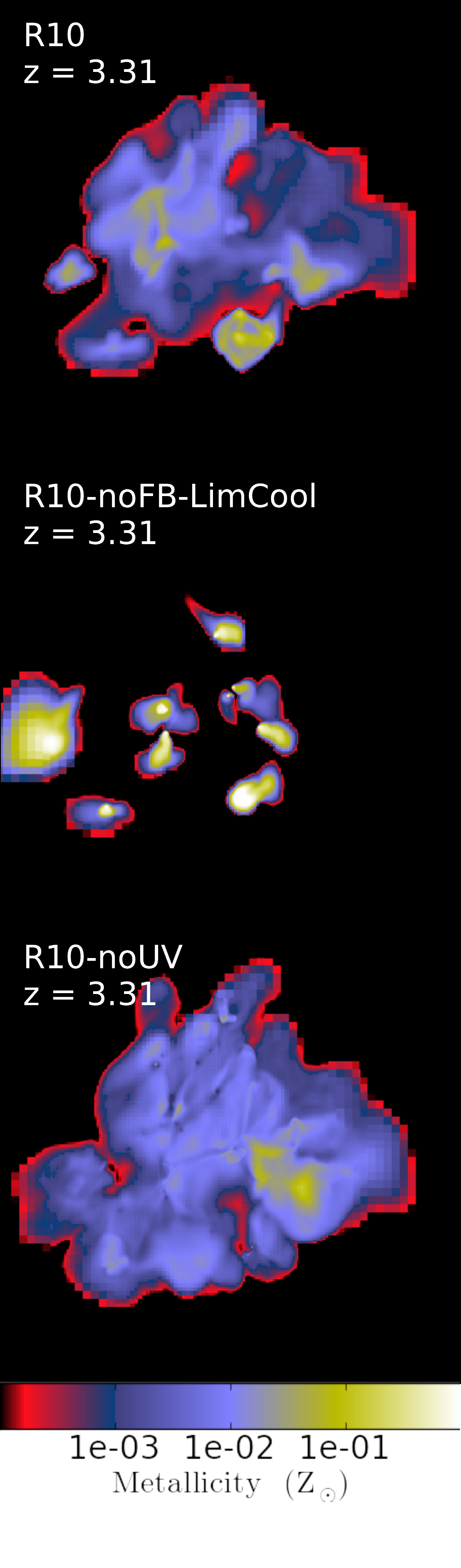}
\caption{Gas density weighted projections of gas metallicity at redshift 3.31 for R10, R10-noFB and R10-noUV.  All panels show projections of 150 kpc cubic volumes that encompass all the progenitor haloes at this redshift.}
\label{fig:metal_pollution}
\end{figure}

Figure~\ref{fig:metal_pollution} shows projections of gas metallicity of a large volume that encompasses the dwarf group at $z=3.3$, which eventually forms the final dwarf halo.  It clearly shows that in simulations with thermal supernova feedback, metals are ejected to much greater distances and more completely fill the volume.  The simulation without thermal feedback, R10-noFB-LimCool, has significantly less IGM metal pollution.  This is despite the fact that the total gas fractions of haloes in R10-noFB-LimCool are quite similar to those seen in R10 (Fig.~\ref{fig:zero_feedback}).  The gas fractions of haloes in R10-noUV are quite elevated at this redshift as compared to haloes in R10 (Fig.~\ref{fig:evo_uvbg}).  Despite this, the extent of metal pollution is relatively similar between the two runs with supernova feedback.  Therefore, we conclude that supernova feedback is the primary mechanism for metal ejection in our model.

Increased enrichment of dark haloes after reionization as shown in Figure~\ref{fig:peak_density_cell_properties} is likely quite complicated and due to a combination of effects.  Since supernova feedback is the mechanism for metal ejection from luminous haloes in our model, the enrichment of dark haloes depends on the proximity of dark haloes to luminous haloes and the speeds of supernova enriched winds \citep[e.g.][]{cenbryan01}.  However, the increased degree of enrichment occurs after the cessation of photoevaporative winds due to reionization.  The large drop in gas fraction seen in Figure~\ref{fig:baryon_properties} indicates the effect of these photoevaporative winds on halo gas.  This decline happens in both R10 and R10-earlyUV just before the increase in metallicity of dark haloes.  It is likely more difficult for supernova driven winds to penetrate haloes undergoing strong photoevaporative winds.  This issue may be an interesting topic for future study.

As we have discussed, stellar populations of dSph galaxies are extremely metal poor and this does not appear to be simply due to their low star formation rates.  \citet{kirby11b} have demonstrated that dSphs in the Milky Way eject almost all of their metals.  Dwarf spheroidals are also observed to be quite gas poor \citep{mateo98}.  As we have discussed, environment most likely plays a key role in the gas content of Milky Way dwarfs.  However, our simulations at least demonstrate that the main mechanism for gas expulsion need not be the same as for metal ejection.  This picture is consistent with the simulations of \citet{maclow99} and \citet{fragile03}, who found, using high-resolution, idealized models of dwarf galaxies, that metals were more easily ejected from galaxies than gas.  

The discrepancy in metallicity between our simulated stellar populations and observed systems described in section 4.1.2 indicates that star-forming gas in our model is too metal rich.  This over-enrichment of star-forming gas may be due to insufficient ejection of metals by our supernova feedback model, despite the apparent large scale enrichment of the IGM on the scale of tens of kpc seen in Figures \ref{fig:image_panel_plot} and \ref{fig:evo_uvbg}.  We therefore cannot confidently make predictions for the density of metals in the IGM produced by dwarfs.  It is certainly possible that feedback mechanisms other than supernovae may play an important role in ejecting metals, such as ionization from young stars.  What our models do show, however, is that in systems where the gas fraction is determined by the cosmic UV background, an extra ingredient that preferentially affects metal rich gas is needed to eject metals.

If supernova feedback is primarily responsible for the stellar metallicity of dSphs, then these systems are an excellent test to judge the efficacy of feedback models, given the clear constraints of observations.  Several simulated models produce low metallicities for Milky Way dSph analogues using a variety of approaches, such as idealized, non-cosmological models \citep[e.g.][]{revaz12}; cosmological models focusing on high redshift evolution \citep{ricotti05, tassis12, wise12b}; coarsely resolved cosmological models of satellite systems of Milky Way type spiral galaxies \citep{okamoto10,sawala12}; and highly resolved cosmological zoom-in simulations like our own \citep{sawala10}.  Most of these studies rely on feedback schemes that compensate for resolutions too low to capture the Sedov phase of supernovae, with the exception of \citet{wise12b}, who have very high resolutions (1 comoving pc) in their high redshift study and find that momentum from feedback, in the form of radiation pressure, is necessary to produce low metallicities.  The model most comparable to our own, which is run down to $z=0$, is that of \citet{sawala10}, who in high resolution simulations ($< 100$ pc) of isolated haloes similar in mass to our own, were able to produce haloes that lie on the observed luminosity-metallicity relation.  It is unclear, however, how natural this result is given that the fraction of metals injected into the hot (or cold) phase can be calibrated in their subgrid model.  In addition, we note that metal mixing remains a serious uncertainty in SPH \citep{booth12}.  Simulations of larger cosmological volumes that contain larger haloes have also found that feedback is important in explaining the evolution of galaxy metallicity and enrichment of the IGM over cosmic time \citep{finlator08}. 

\subsection{Dark matter properties}

Dark matter haloes in cosmological N-body simulations follow a predictable density profile with a steeply-sloped cusp at the centre \citep{navarro97,navarro10}.  Several previous studies have found that baryon physics, particularly supernova feedback, can weaken the central dark matter cusp \citep[e.g.][]{navarro96,gnedin02,governato12}.  Several mechanisms for this transformation have been proposed, such as the response of dark matter to the sudden removal of gas from halo centres \citep[e.g.][]{navarro96,gnedin02}, and the direct transfer of energy from repeated energetic gas outflows to dark matter particles \citep{pontzen12,governato12}.

Measuring the actual dark matter density profiles of galaxies is, of course, quite challenging, but one way is to use the resolved stellar populations in dwarf spheroidal galaxies.  To date, studies comparing the dynamics of dSphs to cosmological N-body simulations have been inconclusive \citep{boylankolchin12,strigari10,veraciro12}.  Density profiles with shallow or steep central slopes have been found to fit the observational data equally well \citep{walker09b}.

\begin{figure}
\includegraphics[width=90mm]{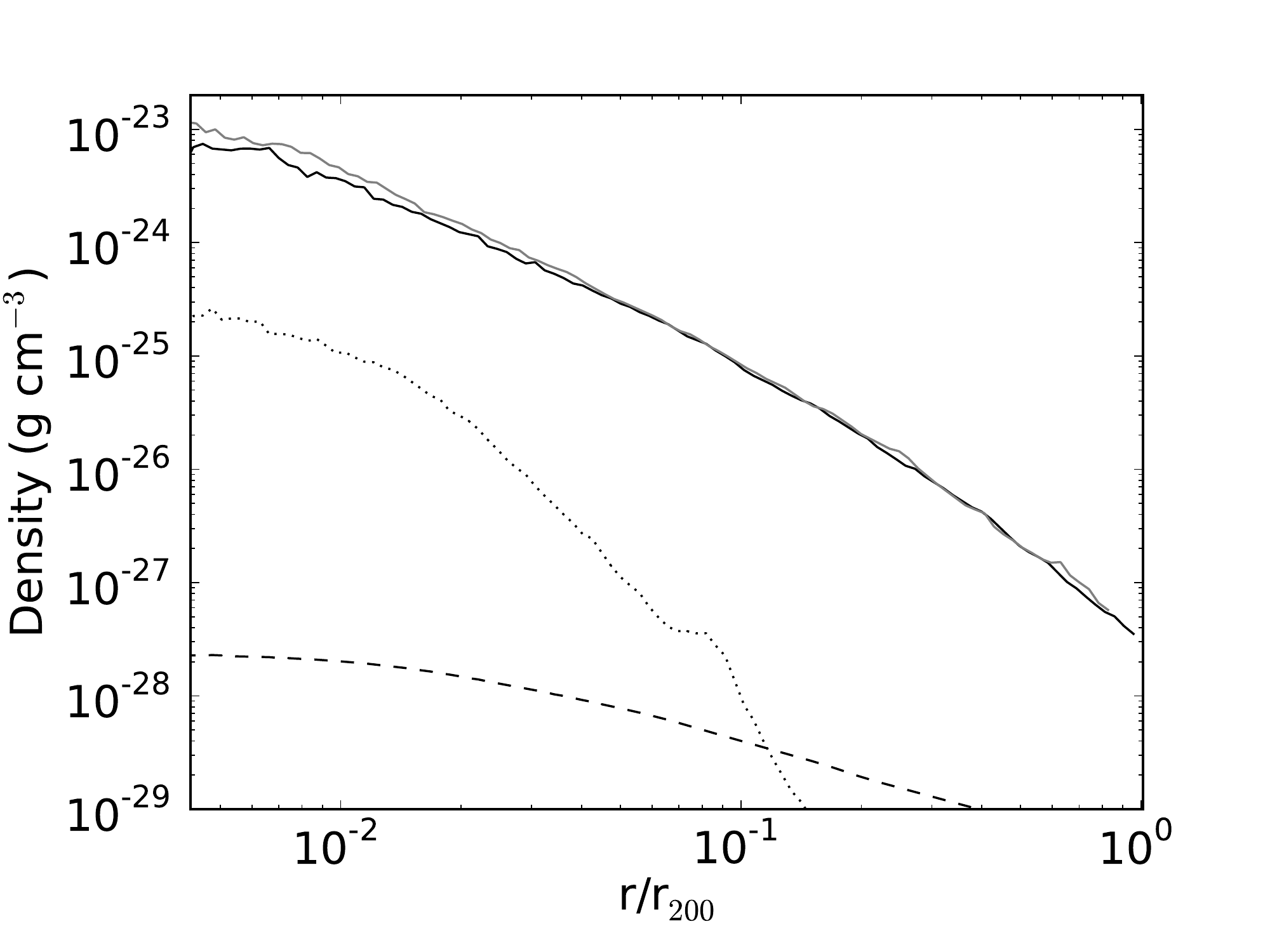}
\caption{Dark matter density profile (solid black line), stellar mass density profile (dotted line) and gas density profile (dashed line) for the final halo at $z=0$ in R10.  The dark matter density profile for the final halo in R10-DM is shown in grey.}
\label{fig:density_profiles}
\end{figure}

Figure \ref{fig:density_profiles} compares the dark matter density profiles for the final halo in R10 and R10-DM, a dark matter only simulation run with the same initial conditions.  We see that the dark matter density profile is almost the same between R10 and R10-DM; the halo in R10 is slightly less dense in the inner regions relative to R10-DM, indicative of a mild heating effect.  We therefore conclude that the impact of baryon physics on the dark matter density was negligible for our halo.  Although not shown in Figure~\ref{fig:density_profiles}, both profiles are well fit by a NFW profile.  The stellar density profile does not follow a NFW profile.  It appears to have a central slope similar to a NFW profile but falls off more quickly beyond 300 pc.  Visual inspection of the final stellar density map shown in the bottom right panel of Figure~\ref{fig:image_panel_plot} also shows that clumpiness and tidal features from late mergers remain, which produces a bump in the stellar density profile.

The one other high-resolution study to examine the density profile of dark matter haloes at this halo mass scale with baryon physics is \citet{governato12}, though using a different hydrodynamical method and stellar feedback model.  They find flattened central density profiles for haloes with virial masses above $10^{10}$ \Msun.  For haloes below this threshold, they find haloes down to our halo mass of $10^9$ \Msun\ have density profiles consistent with NFW profiles, similar to our results.  Haloes in their models do not appear to have consistently flattened central density profiles until the stellar mass rises above $5\times10^6$ \Msun. 

As we have discussed, our supernova driven outflows may not be sufficiently energetic (as evidenced by the metallicity distribution).  It is possible that this failure results in the central density of our halo remaining cuspy, however it is also possible that the low star formation rates of our halo are just insufficient to produce enough energy to have a significant impact on the density profile of our simulated halo.

\subsection{Neglected physics}
\label{sec:neg_phys}
In this section, we briefly discuss physical effects that we do not include in our model.  We do not have an implementation of local UV heating from young stars, either ionizing or dissociating, in our model.  This effect may have an impact on dense gas as young stars will be located within close proximity.  \citet{gnedin10} found that local UV heating sources dominated over the cosmic UV background for both ionizing and Lyman-Werner radiation in haloes with masses between $10^{10}$ and $10^{12}$ \Msun\ at a redshift of three.  We are probing haloes at much lower masses with much lower star formation rates, but it may be that UV heating from young stars can suppress star formation or make supernova feedback more effective by preheating gas.

We do not include any effects of dust in our model.  The formation of H$_2$ on the surface of dust grains is important in gas with metallicities above approximately $10^{-3}$ \Zsun\ \citep{glover03}.  The main effect of H$_2$ in our models is as a coolant in very metal poor gas.  Since the amount of dust in gas with metallicities below $10^{-3}$ \Zsun \ is too small to be important in H$_2$ production, the lack of dust does not significantly affect cooling and therefore star formation in our models.  It does mean that our models do not accurately follow the molecular gas fraction of haloes, which may be an interesting topic of future study.

We do not model early metal-free star formation as a distinct mode of star formation.  \citet{wise12a} presented high-resolution simulations of low mass haloes at high redshift that include the transition between Population III and Population II star formation.  They found that a single Population III supernova can enrich the ISM of a halo to the level where metal line cooling dominates over molecular H$_2$ cooling.

We model supernovae as producing exclusively thermal energy.  Supernovae also produce substantial amounts of radiation, which can not only heat the gas, but can also exert radiation pressure.  The computational cost necessary to do radiative transfer in cosmological calculations is currently too expensive to follow the evolution of haloes down to $z=0$, however several groups have explored its effect at high redshift \citep[e.g.][]{wise12b}.  There are also other effects such as magnetic fields and cosmic rays which we neglect and may potentially be important in galaxies of this size \citep{wadepuhl11}.

\section{Summary}
\label{conclusions}

We have conducted the highest resolution cosmological simulations to date of a 10$^9$ \Msun\ dwarf halo down to a redshift of zero.  Our resolution of 11 comoving pc means we achieve physical resolutions of a few parsecs during the epoch of greatest star formation in our haloes, between redshifts of 6 and 10.  Our canonical runs includes metal cooling, molecular hydrogen formation and cooling, photoionization and photodissociation from a metagalactic background (with a simple prescription for self-shielding), star formation, and a model for supernovae driven energetic feedback.  Our main results are summarized below:
\begin{enumerate}

\item We find that reionization is primarily responsible for setting the total gas fraction of our haloes.  This occurs through photo-evaporative winds, which expel most of the gas envelope in our simulated systems (although not the central, dense core).  Once the gas is ejected, our halo is unable to efficiently re-accrete new baryons, and so the gas fraction remains suppressed to $z=0$.

\item The dense gas, from which star particles form, appears to be mostly regulated by feedback from star formation, although UV heating contributes.

\item The timing of reionization has a major impact on the final stellar mass of our halo, by interrupting the cooling of gas onto lower mass progenitor haloes.  Depending on the patchiness of reionization, an early or delayed epoch of ionization may result in an order of magnitude change to the resulting stellar mass

\item Our final halo has a mass-to-light ratio consistent with Milky Way dSphs.  The systems our models most resemble are the ultrafaint dSphs found in SDSS, although, our simulated halo is located in a much lower density environment than the observed ultrafaints.  The star particles in our models are generally too metal rich compared to Milky Way dSphs with similar luminosities, indicating a problem with our feedback model's ability to eject metals.  Finally, our isolated halo is extremely gas poor at $z=0$, in contrast to many dwarfs in the Local Group and the local volume in low density environments but in agreement with many of the ultrafaint dSphs. 

\item While reionization is the main mechanism responsible for expelling gas from progenitor haloes, supernova feedback is mostly responsible for expelling metals.

\item H$_2$ line cooling is a crucial and necessary ingredient to start star formation in these low mass haloes.  Atomic line cooling in the absence of metals is insufficient by itself to cool gas to the densities required for star formation in such shallow potential wells.  Without a first generation of star formation to pollute the ISM with metals, there is no coolant sufficiently effective to cool gas at subsequent epochs. 

\item We find that the maximum star formation rate and overall duration of star formation is sensitive to the spatial resolution of the calculation.  Our models appear to have not yet converged to a star formation history for our halo, but we argue that the final stellar masses we find for our simulated halo are upper limits.  

\item Our final halo, which has a mass of $10^9$ \Msun\ at $z=0$, forms hierarchically from several progenitor haloes, many of which are star-forming.  The stellar population in the final halo was therefore built up from star formation episodes in these separate haloes.

\end{enumerate}

We conclude that while we do reproduce many of the properties of dwarf systems, such as high mass-to-light ratios and low luminosities, more work is required to understand how stellar feedback works to regulate star formation and metal ejection in low mass dwarf galaxies, and hence the properties of stellar populations in dwarf haloes.  The feedback scheme we adopted was deliberately simple, and not tuned to reproduce observations.  Future work will involve more realistic modeling of supernovae, a model of local sources of ionization, an expansion of the halo we simulate to larger masses and potentially haloes in different environments.

\section{Acknowledgements}
This work was supported by National Science Foundation (NSF) grant AST-0806558 and NASA grant NNX12AH41G.  We also acknowledge support from NSF grants AST-0547823, AST-0908390, and AST-1008134, as well as computational resources from NASA, NSF XSEDE, TACC and Columbia University's Hotfoot cluster.  BDS was supported in part by NASA through grant NNX09AD80G and by the NSF through grant AST-0908819.  M-MML also acknowledges partial support from NSF grant AST11-09395.

CMS would like to thank Mary Putman and David Schiminovich for their support and guidance in the development of this work.  She would also like to thank Jacqueline Van Gorkom for her feedback on our manuscript.

\end{document}